\documentclass[iop]{emulateapj}
\usepackage{natbib}

\usepackage{xspace}
\usepackage{color}
\usepackage{epsfig}
\usepackage{graphicx}
\usepackage{amsmath}
\usepackage{amssymb}
\usepackage{multirow}
\usepackage{rotating}

\def\ssim{\sim\kern-.3em}
\def\aapprox{\kern-.1em\approx\kern-.3em}
\def\lsim{\mathrel{\raise.3ex\hbox{$<$\kern-.75em\lower1ex\hbox{$\sim$}}}}
\def\gsim{\mathrel{\raise.3ex\hbox{$>$\kern-.75em\lower1ex\hbox{$\sim$}}}}
\def\log{\,{\rm log}}

\def\Gyr{\,{\rm Gyr}}

\def\yr{\,{\rm yr}}

\def\Mpc{\,{\rm Mpc}}

\def\dex{{\,\rm dex}}
\def\Ghz{{\,\rm GHz}}

\def\cmm2{{\,\rm cm^{-2}}}
\def\um{{\micron}}
\def\Ha{{\rm H\alpha}}
\def\Hb{{\rm H\beta}}
\def\OIII{{\rm OIII}}
\def\NII{{\rm NII}}
\def\Dage{{\Delta_{\rm age}}}
\def\zerr{{z_{\rm err}}}
\newcommand{\tform}{t_{\rm form}}
\newcommand{\tlb}{t_{\rm lb}}

\newcommand{\SFR}{{\rm SFR}}

\newcommand{\sSFR}{{\rm sSFR}}
\newcommand{\BT}{{\rm B/T}}

\defcitealias{Weisz2011}{W11}
\defcitealias{Karim2011}{K11}
\defcitealias{Oliver2010}{O10}
\defcitealias{Bruzual2003}{BC03}
\defcitealias{Maraston2005}{M05}

\begin{document}

\title{On the Last 10 Billion Years of Stellar Mass Growth in Star-Forming Galaxies}

\author{Samuel N. Leitner}
\affil{Department of Astronomy \& Astrophysics, The
  University of Chicago, Chicago, IL 60637 USA} 
\affil{Kavli Institute for Cosmological Physics and Enrico
  Fermi Institute, The University of Chicago, Chicago, IL 60637 USA} 

\smallskip
\begin{abstract}
The star formation rate - stellar mass relation (SFR-$M_*$) and its evolution (i.e., the SFR main sequence) describes the growth rate of galaxies of a given stellar mass and at a given redshift. Assuming that present-day star-forming galaxies were always star-forming in the past, these growth rate observations can be integrated to calculate average star formation histories (SFHs). Using this Main Sequence Integration (MSI) approach, we trace present-day massive star-forming galaxies back to when they were $10-20\%$ of their current stellar mass. The integration is robust throughout those epochs: the SFR data underpinning our calculations is consistent with the evolution of stellar mass density in this regime. Analytic approximations to these SFHs are provided. Integration-based results reaffirm previous suggestions that current star-forming galaxies formed virtually all of their stellar mass at $z<2$. It follows that massive galaxies observed at $z>2$ are not the typical progenitors of star-forming galaxies today.

We also check MSI-based SFHs against those inferred from analysis of the fossil record -- from spectral energy distributions (SEDs) of star-forming galaxies in the Sloan Digital Sky Survey, and color magnitude diagrams (CMDs) of resolved stars in dwarf irregular galaxies. Once stellar population age uncertainties are accounted for, the main sequence is in excellent agreement with SED-based SFHs (from \texttt{VESPA}). Extrapolating SFR main sequence observations to dwarf galaxies, we find differences between MSI results and SFHs from CMD analysis of ACS Nearby Galaxy Survey Treasury and Local Group galaxies. Resolved dwarfs appear to grow much slower than main sequence trends imply, and also slower than slightly higher mass SED-analyzed galaxies. This difference may signal problems with SFH determinations, but it may also signal a shift in star formation trends at the lowest stellar masses.\end{abstract}

\maketitle
\smallskip

\section{Introduction}

A history of star formation in star-forming galaxies (SFGs) is central
to the subject of galaxy evolution. To an extent, this history can be
extracted from SFGs directly, by fitting stellar population models to
either resolved stars in color magnitude diagrams (CMDs) or to their
spectral energy distributions \citep[SEDs; e.g., ][]{Heavens2000,
Asari2007, Mateus2006, Ocvirk2006, Tojeiro2007, CidFernandes2005,
Panter2007, Williams2011}. Unfortunately, an imperfect knowledge of
stellar evolution, its slow pace in old populations and typical noise
in observations, all conspire to produce broad uncertainty in ages
that are estimated from this ``fossil record'' analysis. Average star
formation histories (SFHs) can also be inferred from the tight
star formation rate (SFR) - $M_\ast$ relation because it describes a tight main sequence of
star formation that connects SFG growth, stellar mass, and redshift.
If present-day SFGs were always SFGs -- always part of this main
sequence in their past -- then these growth rate observations can be
integrated to calculate average SFHs.

Data on the SFR main sequence is ripe for such exploration as
multi-epoch, mass-binned SFR measurements, done through panchromatic
surveys, are rapidly expanding our view of galaxy growth
\citep[e.g.,][]{Oliver2010, Noeske2007st, Elbaz2011, Elbaz2007,
Magdis2010, Pannella2009, Ilbert2010, Karim2011, Feulner2005,
Bauer2005,Bauer2011, Drory2008, Damen2009a, Cowie2008, Zheng2007,
PerezGonzalez2005, Rodighiero2010lfunc, Dunne2009, Wuyts2011}. Surveys
now report representative samples of galaxies that are responsible for
over $70\%$ of star formation up to $z\approx2$
\citep[e.g.,][]{Karim2011,Wuyts2011}. Moreover, the data suggests a
tight SFR main sequence at all masses and redshifts
probed\footnote{\citet{Cowie2008} should be noted as an outlier in
this regard.}. At $z\approx0.1$, the SFR-$M_\ast$ relation is detected
over more than four decades in stellar mass, from at least $10^7M_\odot$ to
$10^{11}M_\odot$ \citep[][]{Wyder2007,
Brinchmann2004,Salim2007,Elbaz2007,Lee2011} with approximately
lognormal $1\sigma\lsim0.3\dex$ scatter.
The small scatter in the relation
persists at $z\approx1$ \citep{Noeske2007data, Elbaz2007}, $z\approx2$
\citep{Daddi2007,Rodighiero2011}, through $z\lsim3$ \citep{Elbaz2011} and out to the earliest epochs at $4<z<7$ \citep[][where the relation may
widen]{Gonzalez2010}.

If a SFG was always a SFG (i.e., formed stars throughout its history),
then the SFR of that SFG's progenitors must appear in the scatter of
the SFR main sequence at a given stellar mass and redshift. That SFR
is the main component governing a SFG's mass growth;
thus, starting with a present-day stellar mass, $M_{\ast0}$, a
galaxy's growth rate from the main sequence ($\dot{M}_\ast$) can be integrated in lookback time
to determine typical $M_{\ast}(z)$ and mass-dependent SFHs (see \S\ref{sec:misfit} for equations). Although,
merging alters how stellar mass is distributed between progenitors, we
demonstrate in \S\ref{sec:mergers} that merging has a negligible
impact on collective stellar mass evolution. If the assumption that
SFGs were always SFGs is correct, then the SFHs derived through this
approach are entirely empirical.

There is good reason to believe that SFGs formed stars throughout
their evolution. At $z\sim0.1$, a combination of characteristics that
are not typically reversible can be used to reliably identify galaxies
as quiescent. Quiescent galaxies have large stellar masses
\citep[e.g.,][]{Brinchmann2004}, and dispersion-dominated kinematics
\citep[e.g,][]{Roberts1994, Kauffmann2003, Wuyts2011structure,
Schiminovich2007}, and exist in regions of high galaxy density
\citep[e.g.,][]{Peng2010, Sobral2011}. At higher redshift,
the quiescent fraction decreases, meaning there are fewer quenched
galaxies that could have been progenitors of present-day SFGs
\citep[e.g.,][]{Bell2007,Pannella2009}. 

If the SFR main sequence reflected on-and-off bursts of star
formation, then a quenched phase could be part of the duty cycle of
SFGs; however, this possibility contradicts abundant evidence pointing
to continuous star formation in observed SFGs. First, from a physical
perspective, the Schmidt-Kennicutt relation \citep{Schmidt1959,
  Kennicutt1998} requires continuous consumption of gas in the absence of
violent disruption to gas disks. Further, \citet{Noeske2007data} noted
that the narrowness of the SFR main sequence itself suggests
continuous evolution, and correlations between short and long term SFR
indicators bear this out \citep[e.g.,][]{Quintero2004,Yang2008,Kriek2011}. The duty cycle in
SFG progenitors is also constrained by the quiescent fraction,
discussed above. Moreover, the distribution of star formation is bimodal, clearly separating into two galaxy populations \citep{Wetzel2011}.
Lastly, a low S\`ersic index structural main sequence
maps onto the SFR main sequence
\citep[see,][]{Wuyts2011structure,Schiminovich2007,Kauffmann2003}. Thus,
for bursts of star formation to be responsible for the SFR main
sequence, they would also need to make galaxy surface brightness
distributions less compact.

Notably, many of these observations hold for the lowest stellar
mass galaxies. In dwarfs, the detection of HI is always associated
with star formation \citep[e.g.,][]{Meurer2006, Lee2011}, the dwarf SFR-$M_\ast$
relation is tight \citep{Lee2011}, dwarfs also show a strong
correlation between star formation over short and long timescales
\citep[e.g.,][]{Lee2009, Meurer2006, Hunter2011}, and less than $10\%$
of isolated (central) dwarfs are quenched \citep[see,][]{Wang2009,
  Haines2007, James2008, Peng2011}.

Still, the assumption that present-day SFGs were always part of the
SFR main sequence may break down where quenching and SFR observations
are not well constrained. For example, at early epoch it is possible
that quenched galaxies rejuvenated star formation following gas rich
mergers. This type of scenario is important to bear in mind, so we
highlight regimes where this Main Sequence Integration (MSI) approach is extrapolated and observations are not
yet representative or well understood (see~\S\ref{sec:sfrlimits} for
further discussion).

A number of other studies have also connected observations of star
formation at different epochs to the history of star formation in
galaxies. \cite{Noeske2007st} was the first to use the SFR main
sequence to infer SFHs, finding that, at $z\lsim1$, galaxies assigned
a formation redshift and exponentially decaying SFHs could match their
SFR main sequence observations \citep{Noeske2007data}. Their
``staged-$\tau$'' model presents one way to account for the data, but
there is no reason to restrict SFGs with that specific
parameterization. A number of groups have made the more basic
assumption that galaxies traced average SFR-$M_\ast$ observations,
thereby producing successful descriptions of the shape of the mass function,
downsizing, merger rates, the buildup of the red sequence, and rising
SFHs in galaxies at $z>3$ \citep[see][]{Drory2008,
Peng2010, Bell2007, Peng2011, Renzini2009, Papovich2011}.
\citet{Leitner2011} evolved SFGs backward starting at $z=0$ and showed
how to incorporate mass loss self-consistently into integration of the
SFR main sequence. The purpose of this paper is to calculate the
quantitative implications of the MSI approach using recent
observations that may have recorded the early stages
of present-day SFG growth. 
  
We will also confront MSI results with fossil record analysis that
determines SFHs for individual galaxies by examining their SEDs and CMDs.
Similarly, both \citet{Walcher2008} and \citet{Wuyts2011} used the SFR
main sequence to calibrate and improve SED-based SFHs, but the focus
here is on understanding the different implications of MSI and the
fossil record. The SED-based results in this paper draw on analysis of
hundreds of thousands of galaxy spectra from the Sloan Digital Sky
Survey \citep[SDSS][]{York2000,Strauss2002} but results are limited
by well-known degeneracies, interpretation of noisy data, and pitfalls
in modeling \citep[e.g.,][]{Ocvirk2006, CidFernandes2007,
  CidFernandes2005, Richards2009}. Using the SFR main sequence as a
basis for understanding these SEDs allows a consistency check on
MSI-based SFHs and an illustration of the effects of age errors on
SED-based SFHs.

CMD-based analysis is uniquely suited to probe early mass growth in
the smallest galaxies. In this regime, MSI is no longer
observationally constrained because the observations are only starting
to probe representative samples of
star formation beyond the local universe at $\lsim10^9 M_\odot$ \citep[e.g.,][]{Wuyts2011structure}. By
extrapolating SFR trends, we can compare MSI to local low mass dwarf
SFHs and see how these smallest galaxies in the local volume fit into
the cosmological trends \citep[e.g., archaeological
downsizing,][]{Cowie1996} in galaxy formation. 

This paper can be divided into two main parts. First, in
\S\ref{sec:sfrsequencedata} and in \S\ref{sec:misfit}, we discuss SFR
data and the SFR main sequence integration method, as well as the implications of
MSI for galaxy growth. Second, in \S\ref{sec:archaeology} we check for
consistency between MSI and analysis of the fossil record.

Specifically, we first describe recent SFR observations
(\S\ref{sec:isfrdata}) and simple fits to SFR data in multiple surveys
(\S\ref{sec:fitdata}) so as to estimate uncertainties. SFR data
limitations are noted (\S\ref{sec:sfrlimits}), both from survey
incompleteness and inconsistency with the growth of stellar mass
density in the universe. We then review how multi-epoch SFR
measurements imply the growth of stellar mass in SFGs
(\S\ref{sec:misfit}). Accurate analytic approximations to SFHs as a
function of stellar mass are provided in
Appendix~\ref{sec:misfit_analytic}. Readers interested in the
importance of scatter or mergers can see \S\ref{sec:complications}.
Early growth in SFGs is described in \S\ref{sec:earlyassembly}, where
results are compared to other semiempirical approaches (Abundance
Matching and Halo Occupation Distributions).

In the second half of the paper (\S\ref{sec:archaeology}), we
compare fossil record-based SFHs with MSI-based SFHs.
\S\ref{sec:archdata} describes the samples we use to span the entire
mass range of SFGs -- from the ACS Nearby Galaxy Survey Treasury
(ANGST) and Local Group (LG) dwarfs analyzed in \citet[][hereafter
\citetalias{Weisz2011}]{Weisz2011}, to SFHs from the versatile
spectral analysis \citep[\texttt{VESPA};][]{Tojeiro2007} database\footnote{See
  \url{http://www-wfau.roe.ac.uk/vespa/} and \citet{Tojeiro2009}.},
which compiles their analysis of the SDSS data release 7
\citep[DR7,][]{Abazajian2009}. We compare MSI to fossil record results
in \S\ref{sec:archresults}, quantifying the importance of deconvolving
errors for interpreting SED-based SFHs (a detailed discussion of age
resolution can be found in Appendix~\ref{sec:ageresolution}).
Finally, in \S\ref{sec:dwarfs}, we highlight the unusual growth history of
dwarf galaxies from the \citetalias{Weisz2011} LG+ANGST sample, using
SDSS SED-based analysis and MSI to place them in a cosmological
context. A discussion of results in the context of simulations and
other observations can be found in \S\ref{sec:discussion}. We
summarize and conclude in \S\ref{sec:conclusions}.

Throughout the paper a flat $\Lambda$CDM cosmology with
$\Omega_0=1-\Omega_\Lambda=0.258$ and $h=0.72$ is used. A
\citet{Chabrier2003} initial mass function (IMF) will be assumed and
all data is converted accordingly.

\section{Measuring the SFR Sequence}\label{sec:sfrsequencedata}

This section describes the measurement and reliability of the SFR
data that we will use to constrain the growth of stellar mass in SFGs.

\subsection{SFR Data}\label{sec:isfrdata}
SFRs in external galaxies are always inferred from tracers of massive
stars. These stars are short-lived and dominate the energy output of
young stellar populations. Their most obvious tracers are at short
wavelengths, including ionizing UV radiation as well as line emission
from elements in the ionized ${\rm H}{\rm II}$ regions (observable in
e.g. ${\rm H}\alpha$ and $[{\rm O}{\rm II}]$) surrounding O and
early-type B-stars \citep{Kennicutt1998, Brinchmann2004}. However,
most of the energy from young stars -- over $80\%$ in SFGs forming
$\gsim2M_\odot\yr^{-1}$ \citep{Buat2010, Takeuchi2010, Bothwell2011}
-- is absorbed by dust and re-radiated in the far infrared (FIR)
spectral window. In fact, UV luminosity from young stars is
subdominant to total IR luminosity ($L_{\rm TIR}$, from $8\um$ to
$1000\um$) in the local universe ($z\sim0.1$) and contributes only
marginally to the total bolometric flux of young stars observed in
SFGs at higher redshifts (see
\citealp{Wuyts2011structure,Takeuchi2005,Bothwell2011}). Long
wavelength observations, which can be more straightforwardly
calibrated to reflect $L_{\rm TIR}$, are therefore reliable tracers of
star formation to early epochs.

The Multi-band Imaging Photometer (MIPS) on the \textit{Spitzer Space
Telescope} \citep{Werner2004}, has provided a crucial window onto dust
SEDs in the mid-IR. \cite{Noeske2007data} used MIPS-$24\um$ to measure the SFR main sequence and
its scatter, thereby constraining staged-$\tau$ models
\citep{Noeske2007st}. More recently, \citet[][hereafter
\citetalias{Oliver2010}]{Oliver2010} analyzed the SFR main sequence over the \textit{Spitzer} Wide-area
InfraRed Extragalactic Legacy Survey \citep[SWIRE][]{Lonsdale2003,
  Lonsdale2004} using the MIPS $70\um$ and $160\um$ bandpasses. The
  lower sensitivity and resolution in these bandpasses necessitated
  stacking galaxies by stellar mass, but the longer wavelength
  observations avoided calibration uncertainties stemming from PAH
  emission features\footnote{Although, \textit{Herschel} FIR observations
  showed that dust properties are uniform across the luminosity function and to $z=1.5$,
  such that old \citep[e.g.,][]{Chary2001} calibrations from
  $L_{24\um}$ to $L_{\rm TIR}$ are accurate to 
  $<0.2\dex$ at $z<1.5$, only significantly overestimating SFRs
  in compact starbursts
  that dominate samples at $z>1.5$ (identified by \citealp{Elbaz2011}; including in \citealp{Elbaz2010,Rodighiero2010,Nordon2010}; see also \citealp{Barro2011}).}
, which redshift to $24\um$ at $z>0.8$ \citep{Fadda2006}.

\citet[][hereafter \citetalias{Karim2011}]{Karim2011} carried out
  the largest probe of the SFR main sequence to date by stacking radio
  observations from the Very Large Array (VLA) in the COSMOS field at
  $1.4\Ghz$ between $0.2<z<3$. Such radio SFR observations are dust
  model independent. However, the radio derived SFRs are only as good
  as their calibration to TIR luminosity: the physical origin of the
  tight empirical correlation between radio and TIR luminosity
  \citep[][]{Bell2003} is not trivially connected to star formation in
  widely varying interstellar medium (ISM) conditions. Fortunately, $z=0$ calibrations seem to
  be accurate\footnote{\textit{Herschel} results show that the ratio of TIR to $1.4\Ghz$ radio emission is
  consistent with a constant value within the errors to $z\approx1.5$
  (see \citealp{Ivison2010b} Figure~5, with \citealp{Jarvis2010}; \citealp{Bell2003}). At $z\leq2$, results
  are consistent with the $z=0$ \citet{Bell2003} calibration to
  $\lsim0.2\dex$ (\citealp[]{Sargent2010a, Sargent2010b,
  Mao2011}; \citealp[][Figure~13]{Bourne2011}).} to $z=2$, but errors
  in the radio-derived SFRs may be dominated by calibration
  uncertainty.

Given \citetalias{Karim2011}'s depth and large statistics, we use that
study for our fiducial dataset. Together, \citetalias{Oliver2010} and
\citetalias{Karim2011} span the results reported in the literature. \citetalias{Oliver2010} 
results are therefore used to estimate analysis
uncertainties as described in the following section.

\subsection{SFR Main Sequence Fits}\label{sec:fitdata}

\begin{figure*}[!t] 
\begin{center}
\vspace{.5cm}
\includegraphics[height=0.55\textwidth]{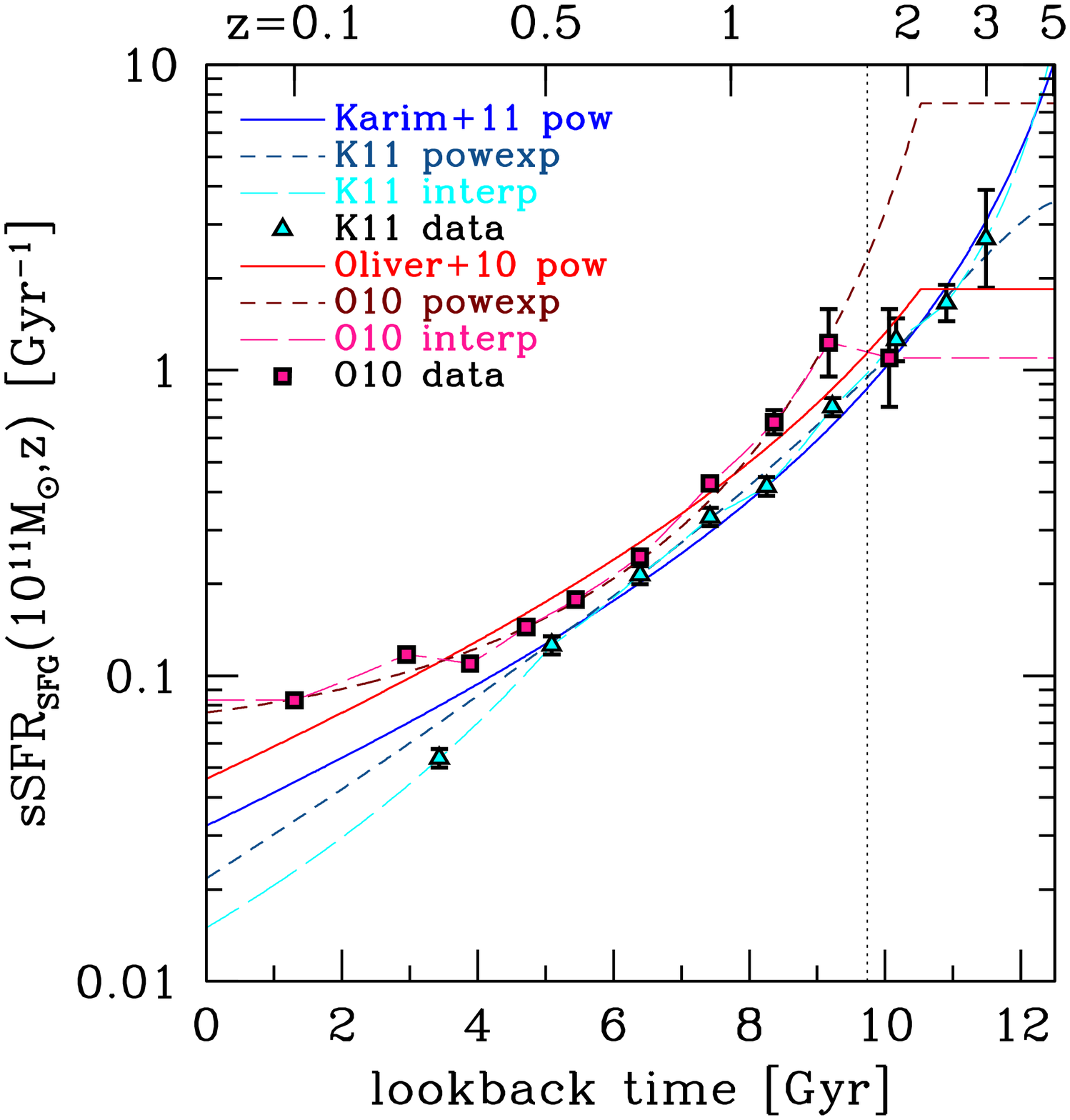}\hspace{0.0ex}
\includegraphics[height=0.55\textwidth]{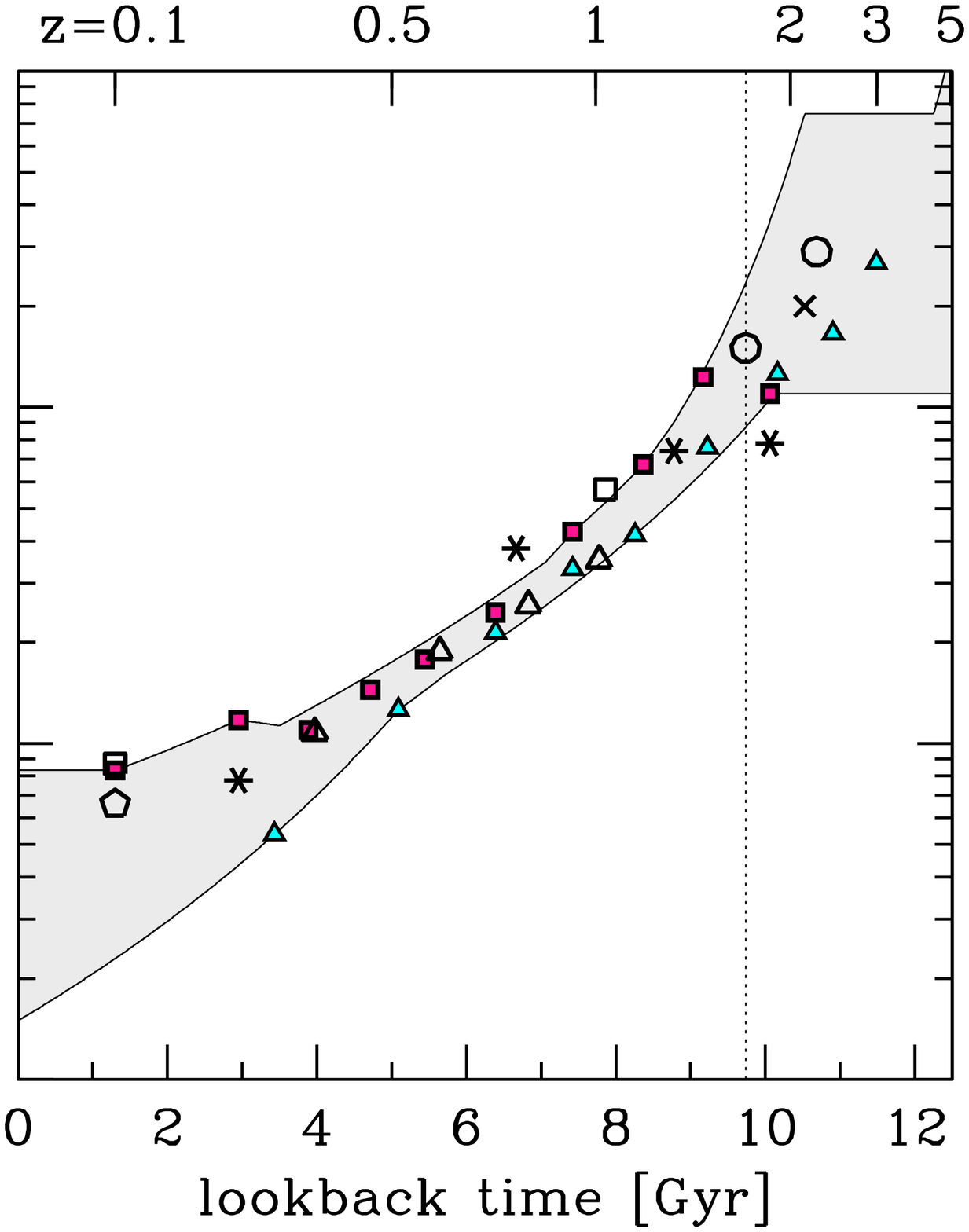}\hspace{0.0ex}\\
\includegraphics[height=0.55\textwidth]{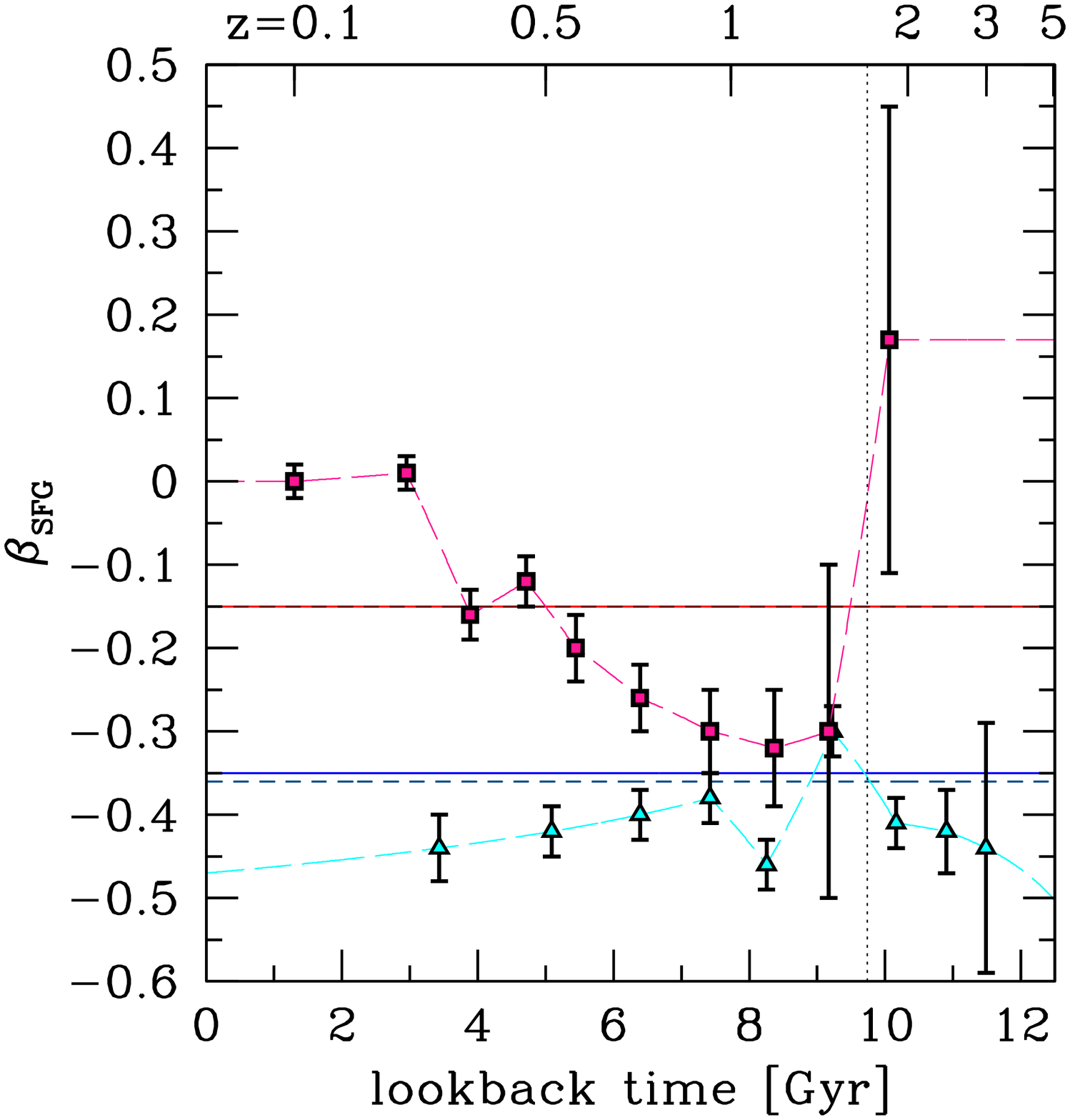}\hspace{0.0ex} 
\includegraphics[height=0.55\textwidth]{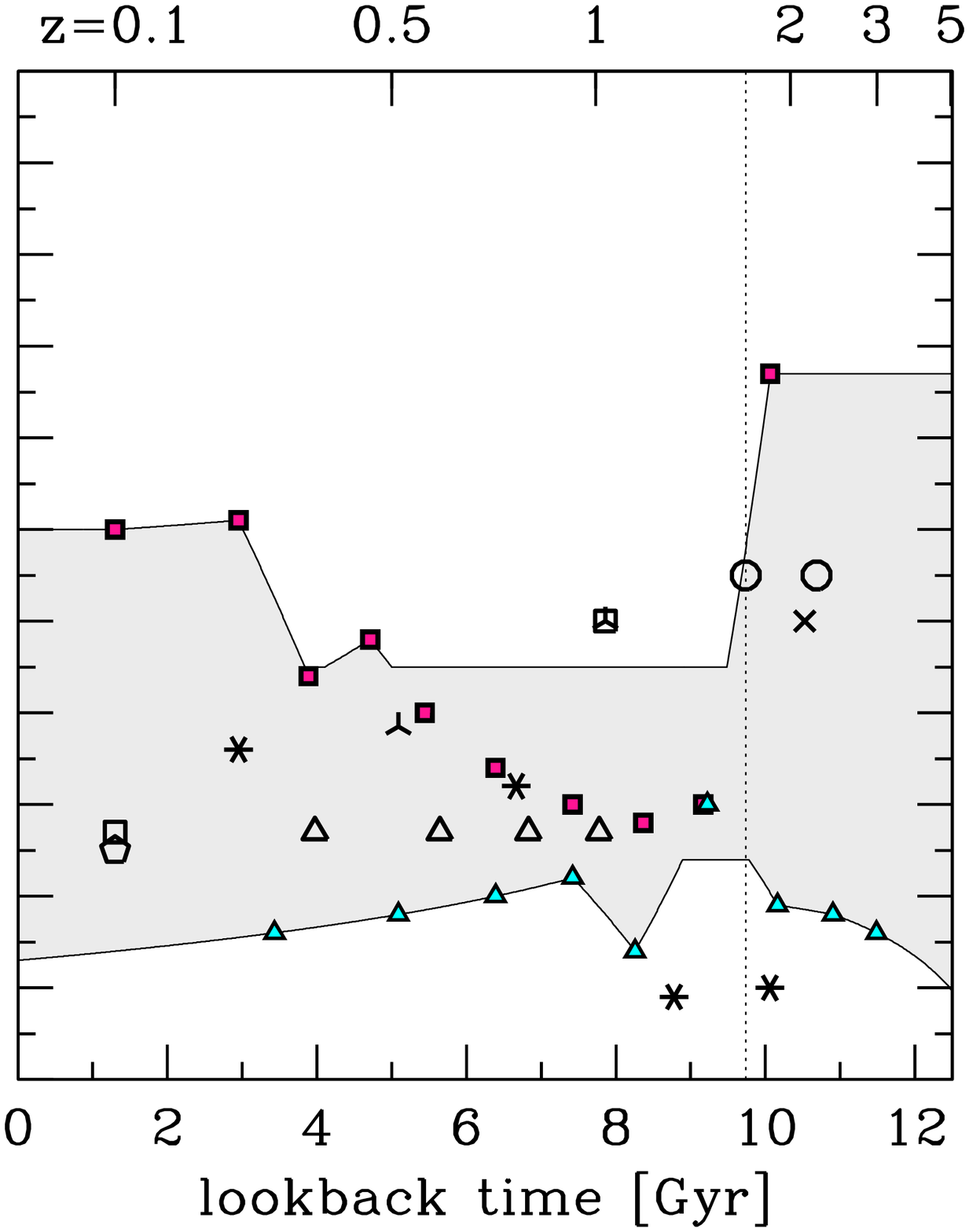}\hspace{0.0ex} 
\end{center}
\caption{ 
{\bf Top panels:} the sSFR of $10^{11}M_{\odot}$ galaxies as a
  function of redshift. 
{\bf Bottom panels:} fits to the  powerlaw slope of the SFR-$M_*$ relation as a function of redshifts.
{\bf Left panels:} lines show fits 
  to \citetalias{Karim2011} (magenta squares) and
  \citetalias{Oliver2010} (cyan triangle) SFR data (see the legend and numbered text for forms). 
{\bf Right panels:} the
  shaded region marks the area enclosed by our fits (lines) from the
  left panels. The vertical dotted lines mark $\zerr$ from
  Figure~\ref{fig:csfhtime}. Points are a compilation of fits to SFR
  main sequence in redshift bins from a number of groups (hollow
  triangles: \citealp{Noeske2007data}; hollow squares:
  \citealp{Elbaz2007}; pentagons: \citealp{Salim2007}; circles:
  \citealp{Pannella2009}; crosses: \citealp{Daddi2007}; three-line
  cross: \citealp{Dunne2009}; asterisks: \citealp{Rodighiero2010}). }
\label{fig:fitdata}
\vspace{10cm}
\end{figure*}

\begin{deluxetable*}{llccccc}[!t]
\tablecolumns{6} 
\tablewidth{0pc} 
\tablecaption{Fit Parameters for $\psi(M_\ast,z)$.\label{table:fitdata}}
\tablehead{\colhead{Data} & \colhead{$\psi$ Form} \hspace*{15 mm} &
\colhead{$A_{11} (\Gyr^{-1})$} & \colhead{$\alpha$} & \colhead{$\beta$} & \colhead{$\eta$} &
\colhead{$\chi^2/n_{\rm dof}$} }
\startdata 
\citetalias{Oliver2010} & pow (Eq.~\ref{eq:powSFRF})       \dotfill  &0.0460 
& 3.36   & $-0.15$ 
& \nodata     & $22.5^*$ \\ 
\citetalias{Oliver2010} & powexp (Eq.~\ref{eq:powexpSFRF}) \dotfill  &0.0759 
& $-3.14$  & $-0.15$ 
& 4.02  & $9^*$ \\ 
\citetalias{Oliver2010} & interpolated \dotfill & \nodata & \nodata & \nodata & \nodata & 0 \\
\citetalias{Karim2011}  & pow (Eq.~\ref{eq:powSFRF}   )    \dotfill  &0.0324 
& 3.45  & $-0.35$ & \nodata     & 1.90 \\ 
\citetalias{Karim2011}  & powexp (Eq.~\ref{eq:powexpSFRF}) \dotfill  &0.0219 
& 5.61   & $-0.36$ & $-0.99$ & 1.15 \\ 
\citetalias{Karim2011} & interpolated \dotfill & \nodata & \nodata & \nodata & \nodata & 0
\enddata 
\tablecomments{Small formal error bars have been omitted, meaningful uncertainties are given by discrepancies between different survey results.\\
*Fits to specific SFR and $\beta_{\rm SFG}$ are based on the redshift-binned data shown in Figure~\ref{fig:fitdata} (not binned by mass).}
\end{deluxetable*}

We integrate fits to the SFR main sequence data to derive MSI-based
SFHs. The fits are done separately to the \citetalias{Karim2011} (radio)
and the \citetalias{Oliver2010} (IR) surveys, thereby providing an estimate of
systematic uncertainties in analysis or calibration that are
otherwise difficult to estimate. They will also be used to extrapolate
measurements to regions of stellar mass and redshift where
SFRs are not well constrained (see~\S\ref{sec:completeness}
for further discussion).

The SFR main sequence is commonly fit to survey data with a power law form in redshift and mass, 
\begin{equation}\label{eq:powSFRF}
\psi(M_\ast,z)=A_{11}\left(\frac{M_\ast}{10^{11}M_\odot}\right)^{\beta+1}(1+z)^{\alpha} .
\end{equation}
Parameters $\alpha$ and $\beta$ are then fit to data in redshift
slices. 
Figure~\ref{fig:fitdata} shows the measured sSFR(z) at fixed
$M_\ast=10^{11}M_\odot$ (which is characterized by $A_{11}(1+z)^{\alpha}$; top left) and
$\beta(z)$ (bottom left) for SFGs from \citetalias{Karim2011}
and \citetalias{Oliver2010}. The right panels of
Figure~\ref{fig:fitdata} show data points from many recent surveys,
demonstrating that measurements typically occupy the region between
\citetalias{Karim2011} and \citetalias{Oliver2010}.

The $\beta(z)$ parameter in Figure~\ref{fig:fitdata} (bottom) shows
the extent to which galaxies of different masses assemble at different
rates. A negative value means more massive galaxies have smaller
specific SFR (${\rm sSFR}\equiv\SFR/M_\ast$) and, therefore, must have
formed a larger fraction of their stellar mass at high redshift. This
is the ``archaeological'' form of downsizing, and is the only form of
downsizing referred to throughout this paper. Some downsizing is
evident from sSFRs -- a process also implied by mass function evolution
\citep[e.g.][]{PerezGonzalez2008, Conroy2009}. The value of $\beta$
and its evolution vary by survey, but \citetalias{Karim2011} showed
that this variation can be largely explained by fluctuations at the
edge of mass ranges, and aggressive color selection \textit{within} the SFR
main sequence\footnote{
Meaningful selection of ``star-forming'' galaxies is not ambiguous. At all masses the galaxy bimodality identifies quenched and star-forming galaxies based on structure \citep[see,][]{Wuyts2011structure} and SFR \citep[e.g.,][]{Wetzel2011}. If the selection criteria accurately reflect the galaxy bimodality (see discussion in \citetalias{Karim2011}), the specific details should not affect the recovered median SFR main sequence. 
}. Despite the use of a wide variety of different methods and calibrations, there is rough agreement 
in sSFR observations, in the sense that trends are larger than the scatter between different studies\footnote{Note error bars are generally underestimated (e.g., left panels of Figure~\ref{fig:fitdata}).}.

In practice, to roughly characterize observations of the SFR main sequence
reported by various studies, we fit three different forms to
the \citetalias{Karim2011} and \citetalias{Oliver2010} SFR data:
\begin{enumerate} 
\item A power law form (``pow'') that follows Eq.~\ref{eq:powSFRF} with constant $\alpha$ and $\beta$. The \citetalias{Karim2011} fit is our fiducial model. The \citetalias{Oliver2010} fit includes a break to a constant $\psi(M_\ast)$ at $z\geq 2$.
\\
\item A power law times exponential form (``powexp'') that follows Eq.~\ref{eq:powexpSFRF} with constant $\alpha$, $\beta$, and $\eta$, 
\begin{equation}\label{eq:powexpSFRF}
   \psi(M_\ast,z)=A_{11}\left(\frac{M}{10^{11}M_\odot}\right)^{\beta+1}(1+z)^{\alpha}e^{\eta z} . 
\end{equation}
We thereby account for curvature in Figure~\ref{fig:fitdata} (left), and
the form allows a formally consistent fit to the \citetalias{Karim2011}
SFG data (reduced-$\chi^2=1.15$ and $n_{\rm dof}=44$).
\\ 
\item A non-parameteric form that linearly interpolates (in $1+z$) between measurements at different redshifts (``interp''). Outside of the data range, constant and linear 
  extrapolations are used for \citetalias{Oliver2010} and \citetalias{Karim2011} respectively.
\\
\end{enumerate}

Parameters of the power law and exponential fits are reported in
Table~\ref{table:fitdata}. For \citetalias{Karim2011} data, we derived
the maximum likelihood $\alpha$, $\beta$ and $\eta$ from their average
stacked SFR results and errors, binned in both mass and redshift, from
their Table~3. For \citetalias{Oliver2010}, the fits are done
separately for the sSFR at $M_\ast=10^{11}M_\odot$ and $\beta$, based
on the points and errors plotted in Figure~\ref{fig:fitdata}. 
Fits are plotted as lines in the left panels of Figure~\ref{fig:fitdata}.

\subsection{Reliability of Results}\label{sec:sfrlimits}

Integration of SFRs to derive SFHs is limited by the availability of
SFR main sequence data and the reliability of that data.

\subsubsection{Completeness Limits and Extrapolations}\label{sec:completeness}

The stellar mass at which SFGs are well represented in a dataset is a ``completeness'' threshold. Our fiducial dataset from \citetalias{Karim2011} reports their $95\%$ statistical completeness threshold -- the stellar mass at which the VLA-COSMOS flux limits and source confusion still allow
$\gsim95\%$ of SFGs to be stacked in
order to measure SFRs at a given redshift. Strictly speaking completeness limits in \citetalias{Karim2011} make it impossible to derive star formation histories for dwarfs, and limits firm results to massive ($\gsim 10^{10}M_\odot$) disks.

However, these limits are conservative, and do not mean that extrapolations are unreliable. On the contrary, there are no
clear indications that the tight SFR-$M_\ast$ relation falls apart at any mass or redshift. Locally,
SFR main sequence trends can be traced across the entire mass function
\citep[][]{Brinchmann2004,Salim2007,Wyder2007}, with 
sSFRs rising toward low masses and
staying high in dwarf galaxies
(e.g.,\citealp[][]{Lee2009UV,James2008,Lee2011}).
Furthermore, the sequence is also in place at low masses at high redshift
\citep{Gonzalez2010}. More recently, \citet{Wuyts2011structure} mapped the sequence to $10^8M_\odot$ and out to $z=2.5$, finding no obvious change in galaxy properties at any mass or redshift. 

Still, data remains limited and
hints of a flattening of sSFRs in low mass galaxies and at high
redshift have been reported (see \citetalias{Karim2011} for further discussion). Regions
where MSI-based star formation histories rely on extrapolated trends are therefore noted
with diagonal stripes in figures below.

\subsubsection{SFRs at $z>1$ and Consistency with Stellar Mass Density}\label{sec:mdotsfr}

A number of studies \citep[e.g.,][]{Hopkins2006, Conroy2009,
Wilkins2008b, Dave2008, Rudnick2006} have reported that the integral
of the cosmic SFR density from $z\gsim 2$ exceeds the growth in
stellar mass density relative to what is observed (modulo mass loss).
Since measurements of stellar mass density are not subject to the same
extrapolation from high mass stars to the bulk of stellar mass, the
discrepancy could imply an evolving IMF
\citep[e.g,][]{Hopkins2006, Wilkins2008a, Wilkins2008b, Fardal2007,
  PerezGonzalez2008, vanDokkum2008,
  Dave2008,Gunawardhana2011,Borch2006}, or a broader failure to
properly model the SEDs of these denser, more active, lower
metallicity, high-redshift galaxies (e.g., \citealp{Nordon2010,Arnouts2007};
although these are increasingly constrained by \textit{Herschel}, see
\citealp{Elbaz2011}).

\begin{figure}[t]
\begin{center}
\includegraphics[width=0.5\textwidth]{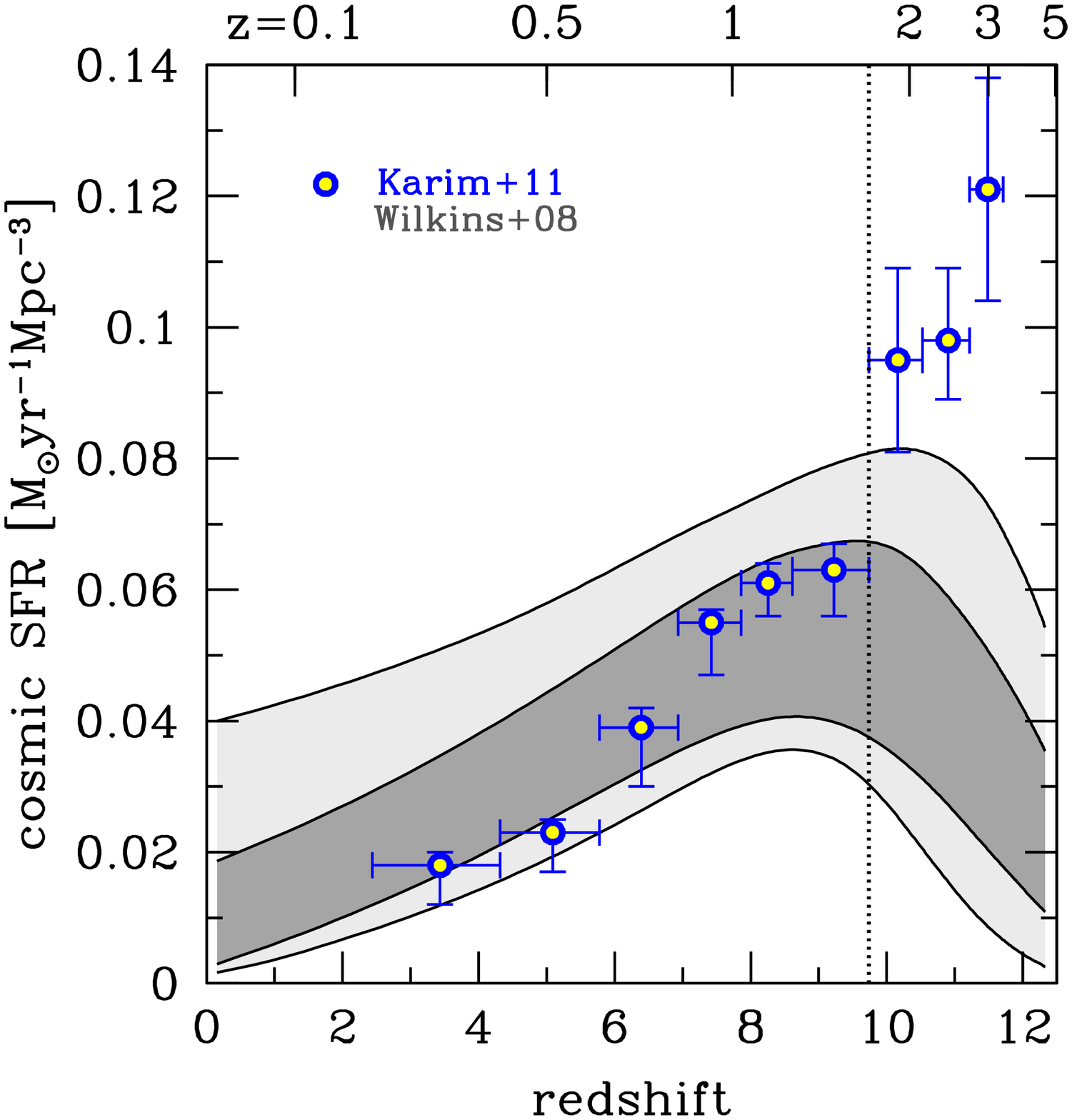}
\end{center}
\caption{Cosmic SFR from the \citetalias{Karim2011} SFR
  sequence data compared to the cosmic SFR expected
  given the growth of stellar mass measured by \citet{Wilkins2008a}.
  The gray region marks the \citet{Wilkins2008a} $1\sigma$ and $3\sigma$
  uncertainty region adjusted from their IMF to a \citet{Chabrier2003} IMF
  by adding $0.045\dex$ based on the difference in bolometric flux
  for a bursting population \citep[calculated in PEGASE.2][]{Fioc1997} .
  The vertical line marks the redshift above which
  SFRs are not well understood ($\zerr\equiv1.6$). }
\label{fig:csfhtime}
\vspace{.5cm}
\end{figure}

In Figure~\ref{fig:csfhtime}, we compare the cosmic SFH from
\citetalias{Karim2011} with predictions for the cosmic SFH based on
the buildup of stellar mass from a compilation by \cite{Wilkins2008a}.
Adjusting for their IMF, the \citetalias{Karim2011} points are consistent
with the 1-$\sigma$ errors on stellar mass growth predictions below
$\zerr\equiv1.6$ (marked by the vertical dashed line). SFR
measurements at $>\zerr$ are discrepant by $>3\sigma$. The stellar
populations that are measured, and the corresponding MSI results,
cannot be reliably interpreted at $z>\zerr$. MSI-based findings at
$z>\zerr$ are therefore noted with dotted shading in figures below.

In the context of past cosmic SFR measurements, \citetalias{Karim2011}
findings are on the low end of the \citet{Hopkins2006} SFR compilation
(see \citetalias{Karim2011} Figure~11), and so indicate that the
\citetalias{Karim2011} SFR main sequence is an outlier. On the other
hand, \citetalias{Karim2011} is consistent with both past IR data and
the growth of stellar mass density in the universe at $z<\zerr$.
Moreover, the total star formation observed by \citetalias{Karim2011}
since $\zerr$ is similar to the total median star formation inferred
by \citet{Wilkins2008a} from the evolution of stellar mass density;
applying a correction for a hypothetical evolving IMF would drive
\citetalias{Karim2011} measurements {\it out} of agreement with SFR data.

While \citetalias{Oliver2010} do not measure a cosmic SFH,
\citet{Rodighiero2010lfunc} included FIR measurements of the same
SWIRE field in their measurements of the cosmic SFH, and
\citetalias{Karim2011} report that these measurements are consistent with
their findings (within large error bars). The \citetalias{Oliver2010}
measurements can therefore be considered valid in a similar
redshift range, but with more stringent completeness limits because of
their shallower SFR observations.

\section{The MSI Approach: from Star Formation Rates to Star Formation Histories}\label{sec:misfit}

With a reliable SFR main sequence in hand at $z\lsim1.6$, observations can
be synthesized through MSI to calculate SFHs, $\Phi(t)$. The
evolution of stellar mass in a given galaxy is described by a growth
term from the observed SFR and a loss term from the galaxy-wide
stellar mass loss rate, 
\begin{equation}\label{eq:mstardot}
\dot{M}_{\ast}(t) = \psi(M_{\ast},z) - \Re(t) .
\end{equation}
Galaxy-wide mass loss is given by an integral over the fractional mass
lost rate ($\dot{f}_{\rm ml}$) from each single age stellar population
(SSP) -- calculated using the flexible stellar populations synthesis
 code \citep[v2.0][or see \citealp{Leitner2011} for fits]{Conroy2010,Conroy2009b}, 
\begin{equation}
\label{eq:grr}
\Re(t) = \int_{t_0}^t \Phi(t') \dot{f}_{\rm ml}(t-t') dt' .
\end{equation}
The SFH of that galaxy is then,
\begin{equation} \label{eq:misfitphi}
\Phi(t) = \psi(M_{\ast}(t),z) ,
\end{equation}
where $\psi$ is a function of the {\it evolving} mass. $\Phi(t)$
and $M_\ast(t)$ can be thought of as the Lagrangian coordinate
tracking a given galaxy, while $\psi$, the SFR main sequence,
is the SFR at a fixed Eulerian coordinate in the space of stellar mass
and redshift.

For our fiducial choice, galaxies start with boundary condition
$M_{\ast0}\equiv M_\ast(z_{\rm obs})$, and trace the median of the SFR main
sequence given by Eq.~\ref{eq:powSFRF} and \citetalias{Karim2011}
parameters (in Table~\ref{table:fitdata}) as they move back in time
and down in stellar mass. Since $\Phi(t)$ is needed to calculate mass
loss, but mass loss is needed to solve Eq.~\ref{eq:mstardot}, a
self-consistent solution for $M_\ast(t)$ requires iteration.
\citet[][]{Leitner2011} described the simple iteration procedure that
is used here to converge on a self-consistent $\Phi(t)$ and $\Re(t)$.

\begin{figure*}[]
\includegraphics[width=0.49\textwidth]{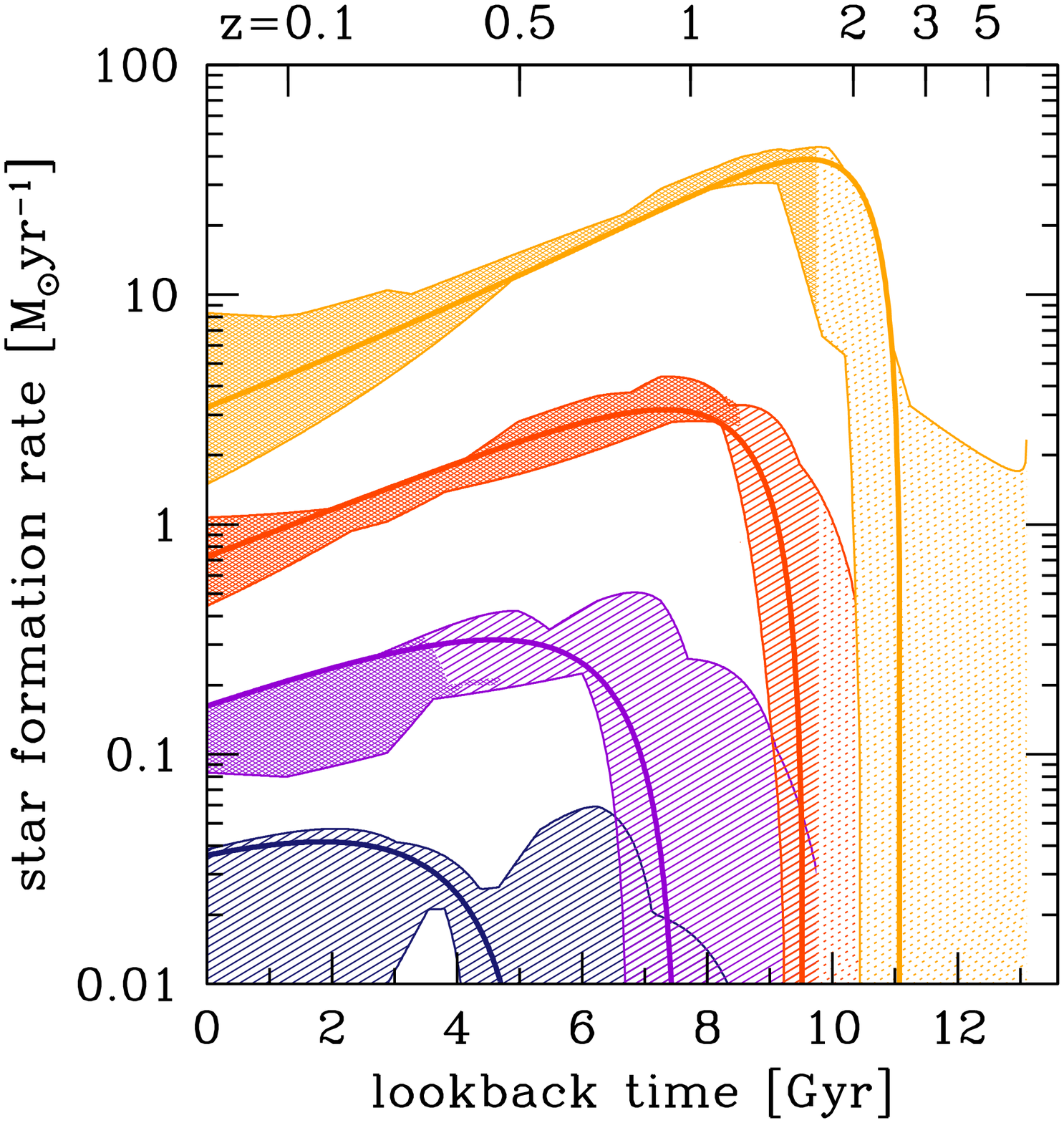}
\includegraphics[width=0.49\textwidth]{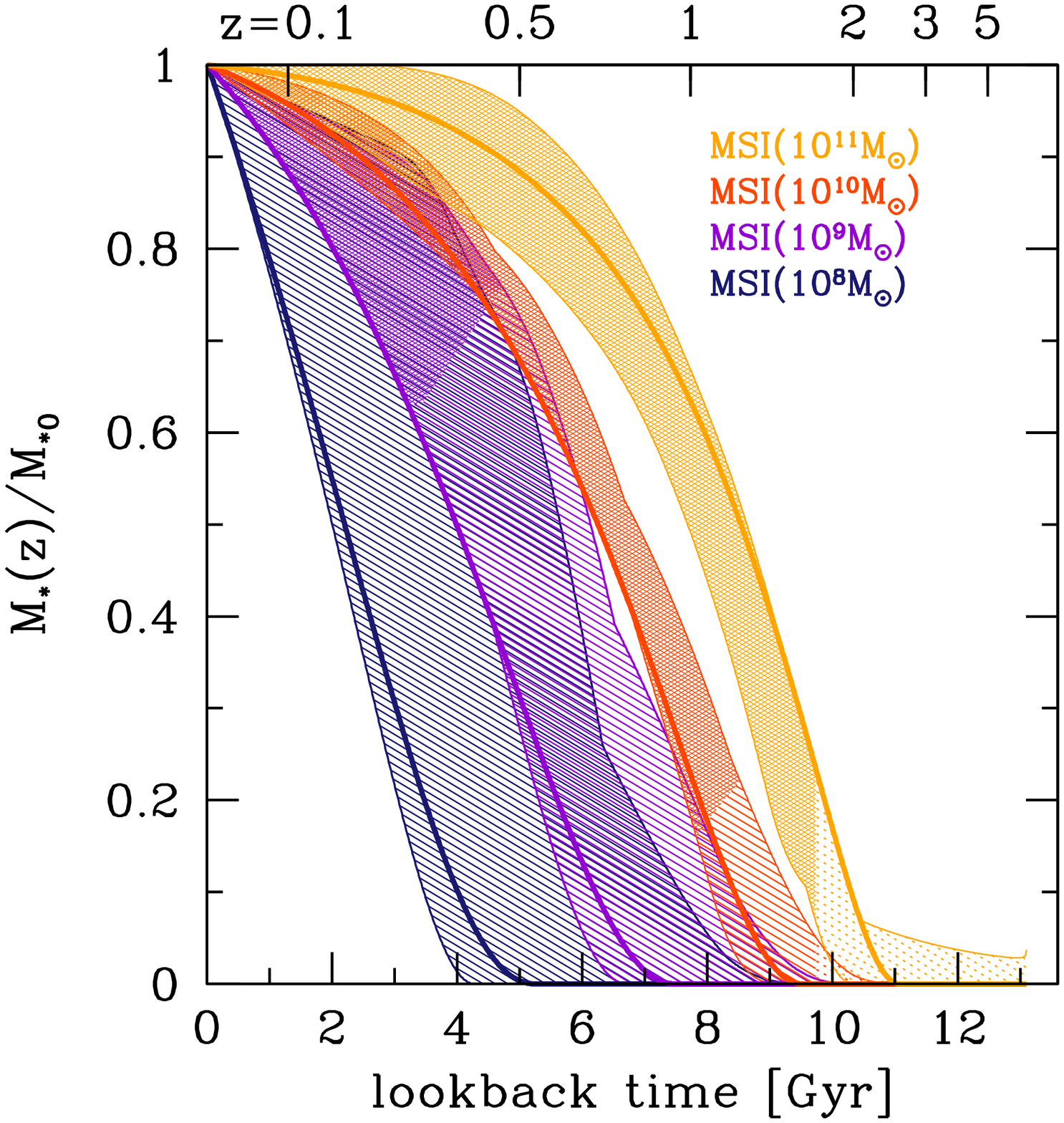}
\caption{ The SFHs ({\bf left}) and mass growth ({\bf right}) from
  MSI, in galaxies of $M_{\ast 0}=10^8M_\odot$, $10^9M_\odot$,
  $10^{10}M_\odot$ and $10^{11}M_\odot$. Shaded regions show the
  variation between SFR main sequence observations and thick lines are
  fiducial power law SFR main sequence fit to \citetalias{Karim2011}
  data. Results are shown where SFR main sequence data is robust and
  complete (cross hatching), extrapolated (diagonal stripes; see
  completeness discussion in \S\ref{sec:completeness}) and compromised
  by observational uncertainties at $z>\zerr$ (dotted). }
\label{fig:allgrowth}
\vspace{.5cm}
\end{figure*}

Figure~\ref{fig:allgrowth} shows mass growth for galaxies including
the spread from variation between different SFR main sequence
observations (described by fits in \S\ref{sec:fitdata}).
Appendix~\ref{sec:misfit_analytic} reports analytic approximations.
MSI-based SFHs decay almost exponentially after early buildup, and are
thus in general agreement with staged-$\tau$ models assumed by
\citet[][]{Noeske2007st}.

\subsection{Assessing MSI Complications}\label{sec:complications}

In this section we address two issues -- mergers and scatter -- that
were ignored by linking galaxies across SFR observations solely using
the median SFR and resulting stellar mass evolution. We find that
these issues are probably less important than variations between
observations of the SFR main sequence,
so readers interested in results can skip to \S\ref{sec:earlyassembly}.

\begin{figure*}[!t]
\vspace{.5cm}
\includegraphics[width=0.49\textwidth]{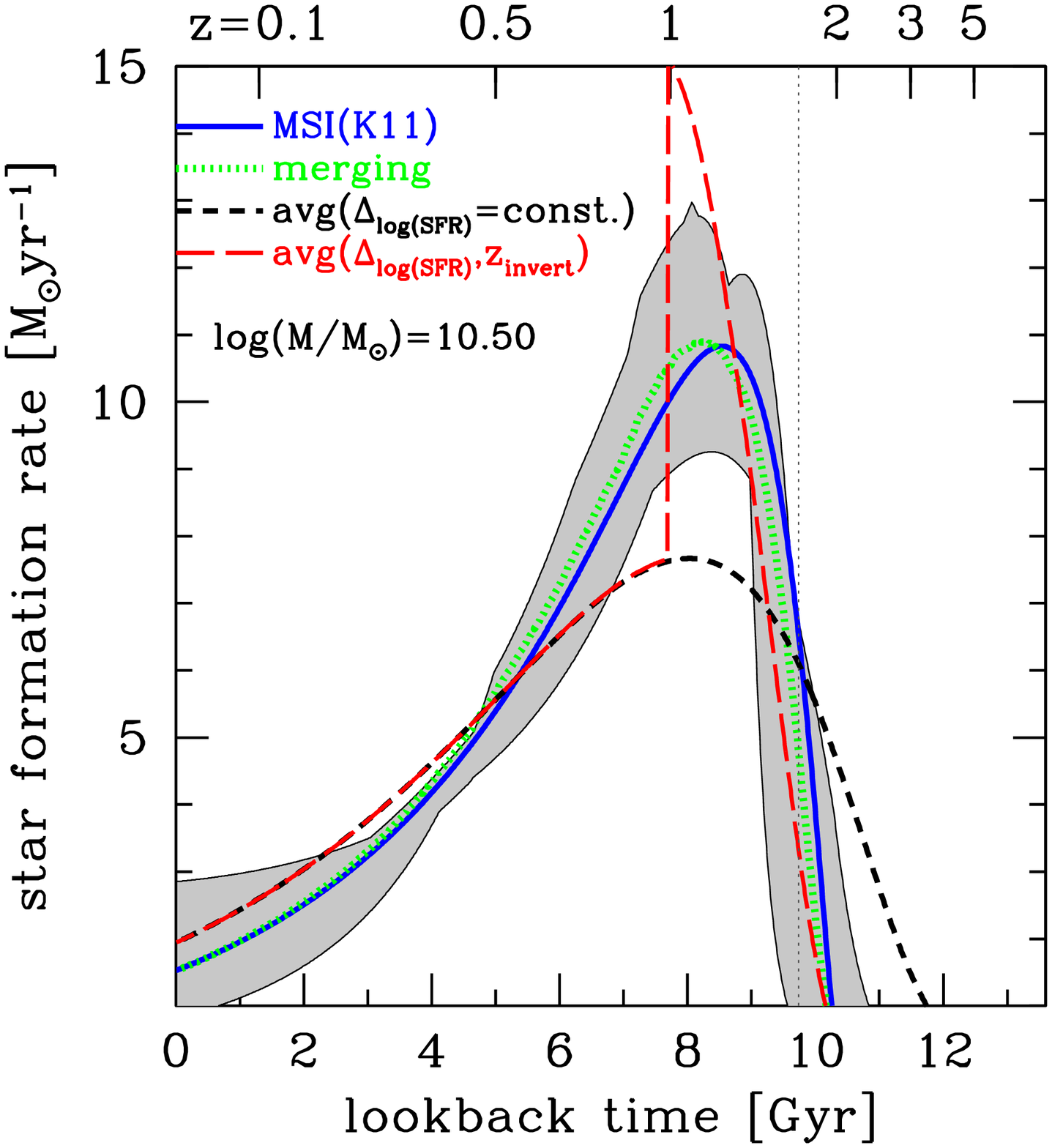}
\includegraphics[width=0.49\textwidth]{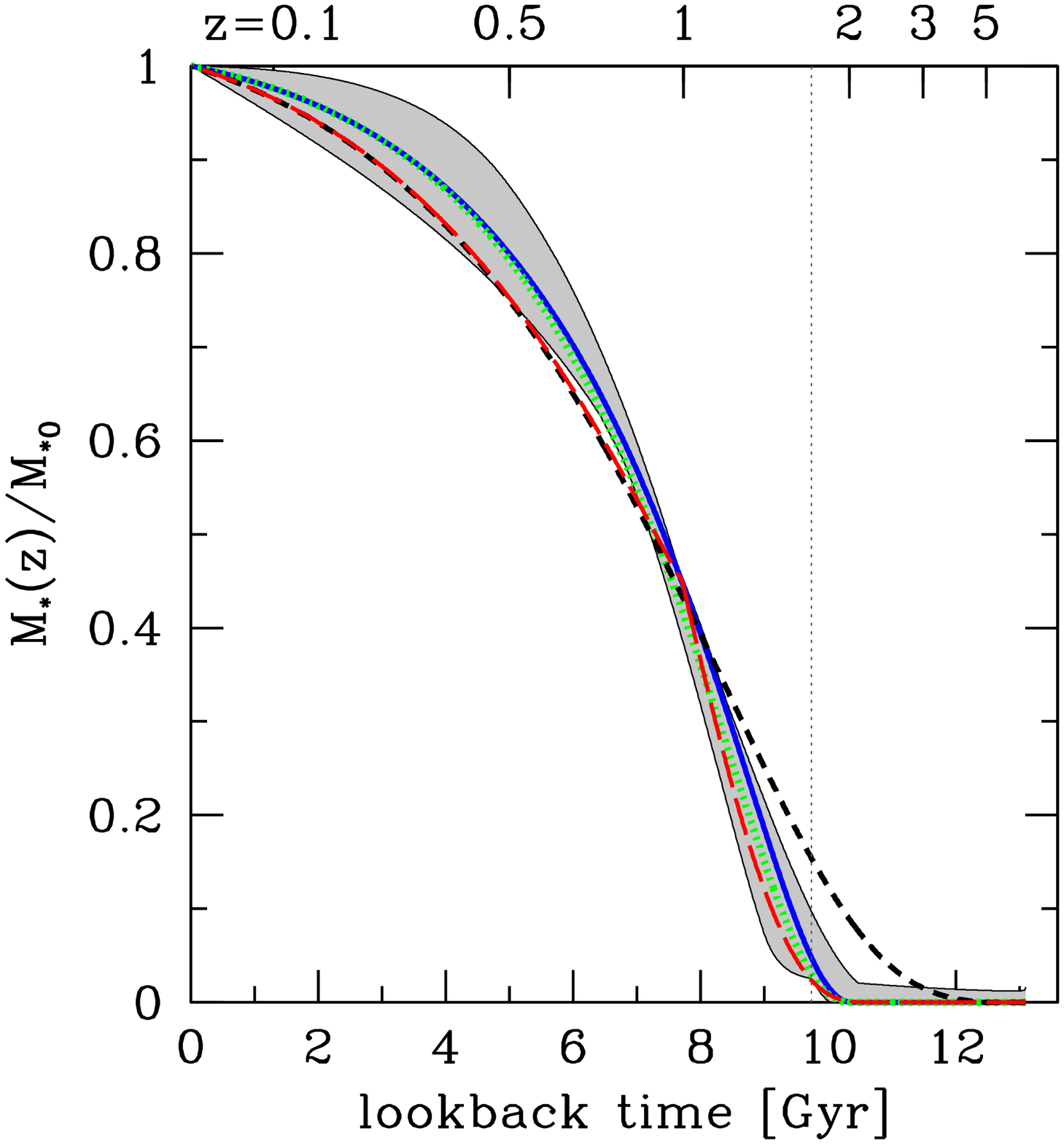}\\
\caption{ Changes to MSI-based SFHs (left) and stellar mass growth
  (right) due to mergers and scatter in SFR-$M_\ast$ for SFGs of
  $M_{\ast0}=10^{10.5}M_\odot$. The gray solid region encompasses
  the spread between MSI based on \citetalias{Karim2011} and
  \citetalias{Oliver2010} data. The blue line shows the fiducial model
  to which other lines should be compared. The green dotted lines
  show averaged MSI merger simulations. The dashed lines show the
  average of Monte Carlo realizations of the lognormal SFR PDF: black
  short dashed shows a constant $\Delta_{\log(\SFR)}$ across time in a
  given galaxy and the red long-dashed line shows a constant
  $\Delta_{\log(\SFR)}$ that inverts at $z=1$. The vertical dotted
  line is at $\zerr$ from Figure~\ref{fig:csfhtime}.
}
  
\label{fig:misfitcomplications}
\vspace{.5cm}
\end{figure*}

\subsubsection{Mergers}
\label{sec:mergers}

MSI is almost insensitive to mergers, both because sSFRs of SFGs are
not strongly dependent on mass and because mergers are not common in
SFGs at $z<2$. Starting from a $M_*(z_{\rm obs})$, galaxies
disassemble according to their growth rate, $\dot{M_\ast}$, with
increasing lookback time. If at some $z_{\rm split}$ there is a merger
of ratio $\mu_{\rm mrat}$, the galaxy gets split into two progenitors,
both of which continue to contribute to the SFH of the final galaxy.
Since sSFRs are mass dependent, the same galaxy grows at a different
rate if it is split into progenitors. The ratio of growth in a split
to unsplit galaxy is
\begin{equation}\label{eq:boost}
\gamma_{\SFR}=\frac{ 1+\mu_{\rm mrat}^{1+\beta}}{ \left(1+\mu_{\rm
mrat}\right)^{1+\beta} } .
\end{equation}
Observations indicate that galaxies downsize ($\beta<0$, see
\S\ref{sec:fitdata}), so low mass galaxies grow faster and
$\gamma_{\SFR}>1$. 
Deviations are maximized for rare equal mass mergers, and
$\gamma_{\SFR}\leq 1.3$.

To confirm that merging is not important to stellar mass growth, we couple MSI with merger rate
estimates, thereby turning MSI into a empirical generator of merger
trees. The merger rate per galaxy, $\nu(M_\ast(t),z,\mu_{\rm mrat})$,
is tabulated from the semiempirical Halo Occupation Distribution
based results of \cite{Hopkins2010a} with code that they
provide\footnote{\url{http://www.cfa.harvard.edu/\~phopkins/Site/mergercalc.html}}.
The (dis-)assembly is then realized with 1,000 Monte Carlo
simulations, where each galaxy and its progenitors evolve according to
both their observed SFR, $\psi(M_\ast,t)$, as well as by splitting
their stellar mass into progenitor systems at each timestep with
probability $\nu(M_\ast(t),z,\mu_{\rm mrat})dt$.\footnote{This is
similar to the method used in \citet{Hopkins2009} to explore bulge
formation.} SFRs and stellar masses are averaged at each timestep and
the resulting SFH of the average galaxy due to merging is plotted in
Figure~\ref{fig:misfitcomplications}, along with the same fiducial
simulation without mergers. Merging results in very slightly younger
SFGs.

Note that this is in no way a physical statement about merger rates or
satellites. It is an empirical observation that splitting a SFG into its typical constituent progenitors should not (on average) substantially change the collective growth history of the galaxy.

\subsubsection{Scatter in SFR-$M_\ast$ and Galaxy Environment}\label{sec:averaging}

In reality, galaxy SFRs scatter around the median SFR-$M_\ast$ value,
with approximately lognormal $0.3\dex$ scatter.
\citep{Elbaz2007,Noeske2007data,Elbaz2011}. 
Taking $\Delta_{\log(\SFR)}$ to be the deviation of a given galaxy
from the median relation in some timestep, there are two extreme
scenarios for populating that scatter: the $\Delta_{\log(\SFR)}$ can
be constant across timesteps, or $\Delta_{\log(\SFR)}$ can be totally
uncorrelated between timesteps. In the correlation coefficient $c_t=0$
case (no covariance), average galaxies form stars at the
average SFR inside of the lognormal scatter (about the median, $\psi$).
 Thus the $c_t$=0 case closely resembles the fiducial case (with the
mean SFR replacing the median). In the case where $c_t=1$, active SFGs
with $\Delta_{\log(\SFR)}>0$ are always active and hence formed a
shorter time ago. Correspondingly, less active SFGs
($\Delta_{\log(\SFR)}<0$) are always less active and therefore formed
at earlier epochs. When averaged, the result is a smeared out SFH as
demonstrated in Figure~\ref{fig:misfitcomplications}.

Observations that parse the breadth of the SFR-$M_\ast$ relation for
SFGs could implicate one of these scenarios. Insofar as galaxy
environment affects accretion history, a second-order correlation of
$\Delta_{\log(\SFR)}$ with galaxy environment would be the most likely
cause of $c_t\neq1$ \citep[see][]{Dutton2010}. A number of groups have
found persistent correlations between environment and SFG {\it
  fraction} \citep[e.g.,][]{Kauffmann2004, Cooper2010, Patel2011,
  Sobral2011, Haines2007}. However, focusing exclusively on SFGs,
there are no conclusive indications that $\Delta_{\log(\SFR)}$ is
significantly driven by environment. At $z\sim 1$, \citet{Sobral2011}
found that SFGs form stars $\sim50\%$ faster in group
environments\footnote{Merging SFGs in clusters form stars rapidly, but
  these are rare, and since such bursts occur in regions with low SFG
  fractions to begin with, they may quench star formation directly (a
  conclusion also inferred from structural data by
  \citealp{Wuyts2011structure} and \citealp{Schiminovich2007}). As part of an extended tail in the
  otherwise lognormal sequence \citep[see,][]{Elbaz2011}, they are likely
  not relevant for the history of present-day SFGs.} relative
to more isolated galaxies of $M_\ast<4\times10^{10}M_\odot$
\citep[consistent with][]{Elbaz2007}. Meanwhile, at 
$z\approx0.1$, findings are mixed: \citet{Peng2010} discerned no difference between
SFRs at fixed mass and varied environment, but \citet{Haines2007}
claimed a significant detection of an {\it inverse}
$\Delta_{\log(\SFR)}$-density relation in low mass SFGs and a marginal
inverse relation spanning $\sim 0.3\dex$ is also apparent in
\citet[][see Figure 15]{Popesso2011} from their low
$0.1\Mpc^{-2}<\Sigma<5\Mpc^{-2}$ to intermediate
$5\Mpc^{-2}<\Sigma<30\Mpc^{-2}$ density environments. Clearly it is
plausible that $c_t=0$, but there is also not enough information to
conclude $c_t\neq0$ -- that either uniform trends in SFG fraction don't
carry over to the SFR-$M_\ast$ relation ($c_t=1$), or that SFGs
encounter an environmental inversion.


To check for the impact of a (hypothetical) environmental
inversion, we re-simulate average MSI SFHs taking the $c_t=1$ model
from the previous section, and impose an inversion in the amplitude
of the scatter for each Monte Carlo simulation at the $z_{\rm
  invert}=1$ time-step. Figure~\ref{fig:misfitcomplications} shows the
average of the realizations. This toy model introduces a sharp feature
in the average SFH because $z<1$ active isolated galaxies have burned
themselves out, leaving a population dominated by relatively passive
systems that are relatively active at $z>z_{\rm invert}$. The
resulting average mass growth and SFHs are similar to the fiducial
model.

Other secondary correlations with $\Delta_{\log(\SFR)}$ may further
constrain the SFHs of SFGs. Properties, such as galaxy clustering,
circular velocity, morphology (e.g., \citealp{Elbaz2007};
\citealp{Wuyts2011structure}), the evolution SFG number counts, and
merging or gas physics, may be used to identify the position of
galaxies in the scatter of the SFR main sequence. Each place
important priors on SFG growth \citep[see,
e.g.,][]{Schiminovich2007,Boissier2010, Bouche2010}. However, the
focus here will be on the implications of consensus SFR observations
in a straightforward empirical sense, rather than on embedding the
integration in a more physics-driven semi-analytic model. These
observations are enough, by themselves, to generate interesting and
easily interpreted constraints on galaxy growth.


\subsection{The Delayed Assembly of Star-Forming Galaxies}
\label{sec:earlyassembly}

Figure~\ref{fig:allgrowth} indicates that measuring the early
stages of present-day SFG evolution does not require observations of the dawn of
galaxy formation. Galaxies observed in the SFR main sequence at high
redshifts (e.g., $z>3$) are mostly quiescent and massive by
$z=0$, while present-day SFGs appear to have grown most of their mass
starting at relatively low redshift ($1<z<2$).

\begin{figure*}[!t]
\begin{center}
\vspace{.5cm}
\includegraphics[width=0.49\textwidth]{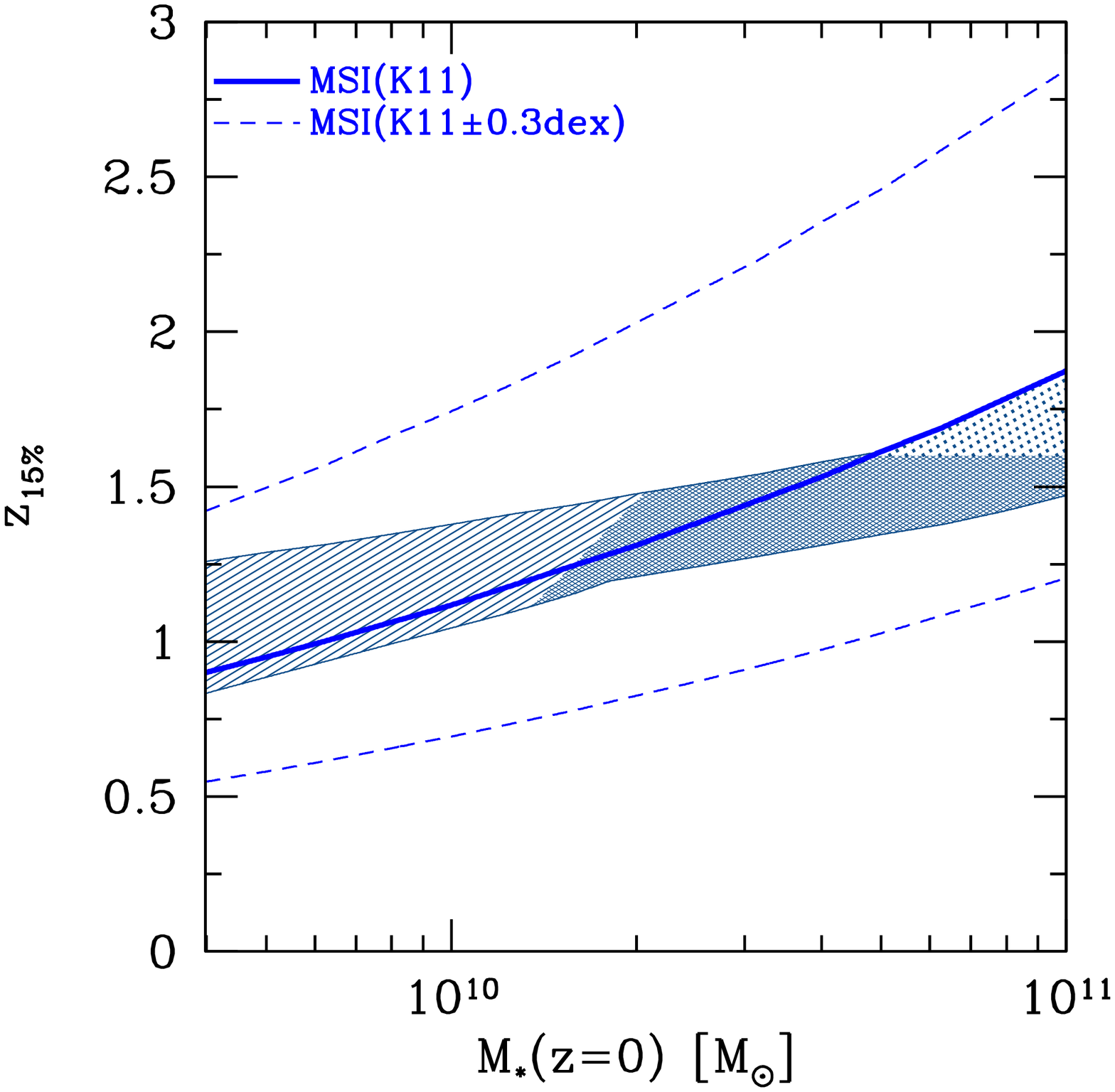}
\includegraphics[width=0.49\textwidth]{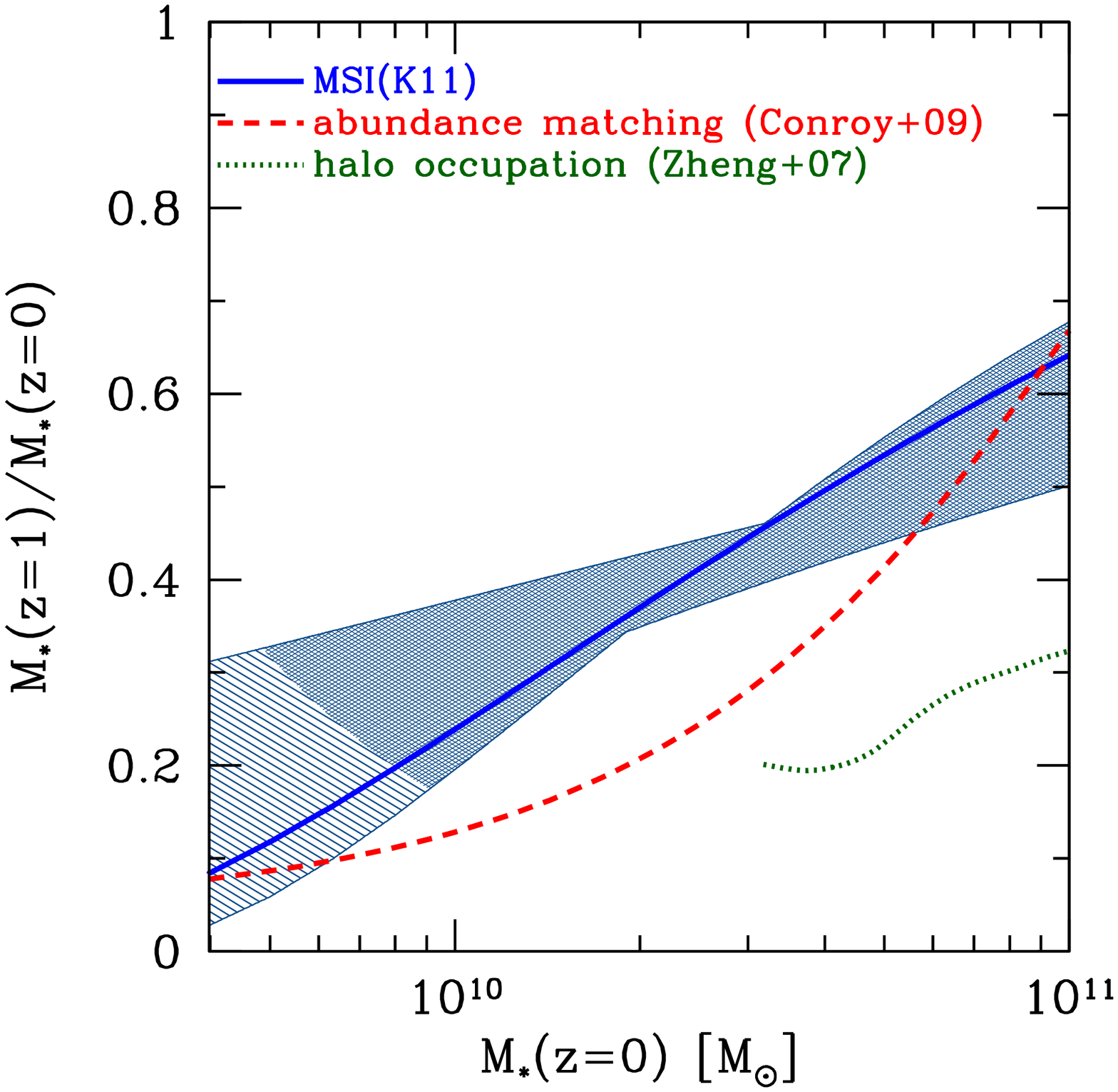}
\end{center}
\caption{ 
Left panel: the redshift by which $15\%$ of stellar mass formed
  in all progenitors of a galaxy from MSI. 
  The shading encompasses variation between
  \citetalias{Karim2011} and \citetalias{Oliver2010} SFR main sequence
  observations. Striping indicates the reliability of the results as
  described in Figure~\ref{fig:allgrowth}. 
The dashed lines show, for reference,
  the effect of large scatter (or error) in $z_{15\%}$ calculated by
  adding $\pm0.3\dex$ to the \citetalias{Karim2011} (thick blue) SFR main sequence.
Right panel: the average
  stellar mass fraction formed by $z=1$ in all progenitors as a
  function of $z=0$ stellar mass from MSI (blue as above), Abundance Matching
  (red dashed), and Halo Occupation modeling (green dotted). }
\label{fig:form}
\vspace{.5cm}
\end{figure*}

\subsubsection{Delayed Growth and Relevance to Bulge Formation }
\label{sec:early}

SFR main sequence measurements from \citetalias{Karim2011} are deep
enough that MSI can follow the growth of stellar mass back to a time
when typical SFGs were almost a tenth of their current size.
Figure~\ref{fig:form} (left) shows the redshift after which MSI
indicates average SFGs formed $15\%$ of their stars ($z_{15\%}$).
The thin lines in the figure show the effect of a factor of two increase
or decrease in the fiducial SFR main sequence; this can be
interpreted as illustrating large systematic uncertainties, or a
maximal estimate of the $68\%$ scatter in SFHs (i.e., in the $c_t=1$
case from \S\ref{sec:averaging}, although ignoring $\zerr$).

We focus on $z_{15\%}$ specifically because, for $10^{10}M_\odot
\lsim M_{\ast0} \lsim 5\times10^{10}$, constraints on $z_{15\%}$ are
still data-driven; $z_{15\%}$ results are affected by neither incompleteness, nor
disagreement with mass function determinations ($\zerr$ from
Figure~\ref{fig:csfhtime}). This regime also bears particular
relevance for disk galaxy structure and bulge formation in simulations
since median observed bulge mass fractions are also $15\%$ (see
\S\ref{sec:discussion_sims} for further discussion). Moreover,
focusing on this mass fraction circumvents complicating factors
related to bulge formation (e.g., rejuvenated star formation from
mergers).

The figure indicates that the peak of star formation in SFGs is
delayed relative to high-redshift ($z\gsim2$) star formers: $85\%$ of stellar mass in many SFGs appears to form
after $z\approx1.5$.
\subsubsection{Model Comparisons at $z=1$} \label{sec:modelcomp} 

Complementary semiempirical methods, based on the
galaxy-dark matter connection and simulations, also expect delayed formation
of SFGs. Figure~\ref{fig:form} (right) shows
the fraction of mass formed in galaxies by $z=1$, including results
from Abundance Matching \citep[][data from their Figure~3]{Conroy2009} and
Halo Occupation analysis \citep[][data from their Figure~7]{Zhengzheng2007}.
These methods paint galaxies onto dark matter halos in simulations at
different epochs according to mass (Abundance Matching) or clustering
(Halo Occupation), and then link galaxies across epochs using average
dark matter accretion rates onto those dark matter halos during the
intervening time. Note that these methods are applied to the galaxy population
as a whole, rather than being restricted to SFGs; however, at $M_\ast
\sim 2\times10^{10}M_\odot$ galaxies transition to being mostly star-forming so the comparison is useful.

Both Abundance Matching and MSI predict lower mass galaxies formed a large
portion of their mass after $z\approx1$, while galaxies of $\sim
10^{11}M_\odot$ form most of their mass at higher redshifts.
Unexpectedly, however, \citet{Conroy2009} find that even less stellar mass
was in place in $z=1$ progenitors than MSI infers -- the addition of
quiescent galaxies should increase the fraction of stars that form
early. The \citet{Zhengzheng2007} Halo Occupation analysis predicts
that even smaller stellar masses were formed before $z=1$ relative to
Abundance Matching. For now, systematic errors in the halo-stellar
mass relation can account for the $0.3\dex$ tension between different
predictions \citep[e.g., from errors on stellar masses and sample
variance, see][]{Behroozi2010,Leauthaud2011}. Indeed,
Figure~12 of \citet{Behroozi2010} explicitly shows that
\citet{Zhengzheng2007} halos host fewer stars than most other
determinations predict at $z=1$. With tighter constraints, differences
from halo-based predictions might be used to infer star formation duty
cycles or the possibility of a quenched phase.


\section{Comparison of MSI with Fossil Record Analysis}
\label{sec:archaeology}

SFHs are more traditionally derived through analysis of a galaxy's
fossil record. This analysis uses either a galaxy's SED or, in the few
local cases where individual stars can be resolved, stellar CMDs.
Given an SED $F_\lambda$, a stellar census $\Phi(t_{\rm age})$ can be
calculated by inverting $F_\lambda=\int\Phi(t_{\rm age})
S_\lambda(t_{\rm age}, Z ) dt_{\rm age}$, where $S_\lambda(t,Z)$ is the
SED produced by a SSP in a population
synthesis model that includes the metallicity ($Z$) dependent stellar
evolution tracks, an IMF, and dust extinction. CMD-based SFHs are
similarly derived, but in this case stellar tracks are fit directly
instead of being summed over. 

Unfortunately, this inverse problem is strongly ill-conditioned; low
levels of noise in $F_\lambda$ can result in widely varying ages for
stellar population, with serious implications for interpretation of
SED-based SFHs \citep[see][]{Moultaka2004, Moultaka2000,
Ocvirk2006, CidFernandes2005, CidFernandes2001}. Moreover, large
statistics can insidiously shrink error bars without improving
recoverable resolution. Still, fossil record analysis provides
important clues to the growth of mass in galaxies, and is singularly
important for measuring the growth of dwarf galaxies where SFR main
sequence data is sparse.

\subsection{ Fossil Record Data: Samples from SDSS and LG+ANGST Dwarfs } 
\label{sec:archdata}

\begin{deluxetable*}{lccccc}[!t]
\tablecolumns{6} 
\tablewidth{0pc} 
\tablecaption{Galaxy Samples \label{table:arch}}
\tablehead{
\colhead{Mass Range $[M_\odot]$}  & 
\colhead{${\rm N}_{\rm gal}$  } &
\colhead{$\langle M_*(z_{\rm obs})[M_\odot]\rangle $} & 
\colhead{$\langle z_{\rm obs}\rangle$} & 
\colhead{$\langle \tlb [\Gyr] \rangle $} & 
\colhead{Source} }
\startdata 
 $M_\ast\lsim10^{8}$    &$32$ &  $4.0\times10^7 $   & $D<4\Mpc$   & $\sim 0$ & LG+ANGST W11 dIrr (CMDs)  \\ 
 $10^8-10^{8.5}$    &$1,115$  &  $2.1\times10^8 $   & $0.014$ &$0.19$ & SDSS$_{\rm SFG}$  (SEDs) \\
 $10^9-10^{9.5}$    &$12,156$ &  $2.1\times10^9 $   & $0.032$ &$0.43$ & SDSS$_{\rm SFG}$  (SEDs) \\
 $10^9-10^{10}$     &$39,390$ &  $5.0\times10^{9}$  & $0.042$ &$0.57$ & SDSS$_{\rm SFG}$  (SEDs) \\
 $10^{10}-10^{10.5}$&$50,071$ &  $2.0\times10^{10}$ & $0.070$ &$0.92$ & SDSS$_{\rm SFG}$  (SEDs) \\
 $10^{10.5}-10^{11}$&$55,780$ &  $5.8\times10^{10}$ & $0.098$ &$1.26$ & SDSS$_{\rm SFG}$  (SEDs) \\
 $10^{11}-10^{11.5}$&$29,187$ &  $1.7\times10^{11}$ & $0.12 $ &$1.61$ & SDSS$_{\rm SFG}$  (SEDs)\\
\enddata 
\end{deluxetable*}

In this section, we briefly describe the fossil record samples and
analysis methods used. Table~\ref{table:arch}
summarizes the relevant properties of the sub-samples from which SFHs
are derived.

SED-based SFHs (for galaxies with $M_\ast>10^8M_\odot$) are drawn from
the \texttt{VESPA} database\footnote{\url{http://www-wfau.roe.ac.uk/vespa/}}
\citep{Tojeiro2007,Tojeiro2009}, which stores SFHs from \texttt{VESPA} applied
to the spectroscopic sample of the of the SDSS DR7. SFGs are selected
from the main galaxy sample
\citep[MGS][]{Strauss2002} using emission line measurements from the
general purpose SDSS pipeline, in order to exclude both quiescent galaxies and
active galactic nuclei(AGNs). SFGs are required to have $\Ha$ emission detected at $>3\sigma$
significance. If $\Ha$, $\Hb$, [$\NII$] ($\lambda 6584$\AA), [$\OIII$]
($\lambda 5007$\AA) are all detected at $>3\sigma$, then we exclude AGN
with a cut in the BPT \citep{Baldwin1981} diagram from
\citet[][Eq.~1]{Kauffmann2003},
$\log\left(\frac{[\OIII]}{\Hb}\right)<0.61\left[\log\left(\frac{[\NII]}{\Ha}
    -0.05 \right)\right]^{-1} +1.3$, and the requirement that
$\NII/\Ha<0.6$; these cuts are in the spirit of
\citet{Brinchmann2004}. No efforts have been made to account for
galaxy selection, but we note that harsher color cuts (e.g.,
$U-R<1.5$) selecting more active SFGs had no effect on our results. Placing a
low S/N threshold (median S/N$>10$ in all bands with unmasked \texttt{VESPA}
$\langle S/N \rangle>20$) also made little difference to SFHs.

The total SFG sample includes about 175,000 galaxies that are split
predominantly into three \texttt{VESPA} measured stellar mass bins of
$10^{9-10}M_\odot$,$10^{10-10.5}M_\odot$ and $10^{10.5-11}M_\odot$.
Two smaller bins, at very low ($10^{8-8.5}M_\odot$) and high
($10^{11-11.5}M_\odot$) mass, are also recorded. \citet{Tojeiro2007}
have shown that \texttt{VESPA} SFHs are consistent with SED analysis
from texttt{MOPED} \citep{Heavens2000}, but \texttt{VESPA} also somewhat mitigates
over-fitting pitfalls by limiting parameters recovered as advocated by
\citet[][]{Ocvirk2006}. The sample therefore represents the best SFH
analysis of an unsurpassed number of galaxy fossil record observations.

For lower mass dwarf galaxies, we use the SFH compilation by
\citetalias{Weisz2011} from the ANGST
\citep[][]{Dalcanton2009} measured by \citet{Weisz2011ANGST},
and from the LG \citep[][Table~1]{Mateo1998}, with CMDs
from \citet{Dolphin2005} and \citet{Holtzman2006}. The sample is
volume limited to $D\lsim4\Mpc$. Individual stars were resolved and
fit with stellar tracks in a CMD. Present-day type and color has
little bearing on past SFH beyond the last $1-2\Gyr$
\citep{Weisz2011ANGST,Weisz2011}, but we will confine our discussion
to the relatively isolated LG+ANGST dwarf irregular (dIrr) sample
\citepalias[][]{Weisz2011} as a low-mass extension of our normal star-forming sample. The
\citetalias{Weisz2011} analysis was performed with a
\citet{Salpeter1955} IMF, so stellar masses are 
multiplied by $-0.25\dex$ to convert to a \citet{Chabrier2003} IMF.
Their detection limits make IMF differences unimportant for cumulative
star formation plots (see \citetalias{Weisz2011}).

\subsection{ Uncertainties in Fossil Record Analysis} \label{sec:noisyarch}

To account for age errors, we note that SSPs change 
logarithmically with age intervals and therefore
convolve MSI results with a Gaussian filter in log-age. In order
to choose a filter width, we note that the SFR weighted average bin
size in \texttt{VESPA} is $\sim0.4\dex$ and covariance between adjacent bins further reduces resolution.
For much higher S/N SEDs, \citet[][]{Ocvirk2006} found that
disentangling confounding degeneracies, coupled with meager SSP
differences, lead to age resolution (full width at half-maximum , $\Dage$) limits of $\gsim0.8\dex$. Below we 
present results smoothed with $\Dage=1.0\dex$, but findings are 
unchanged at $>10^9M_\odot$ assuming any $0.5\dex<\Dage<1.0\dex$ (see
Appendix~\ref{sec:ageresolution} for further discussion). 

\citetalias{Weisz2011} analysis reports broad age bins, such that
covariance in simulated data result in uncertainties that are within
measurement errors; but, again, this covariance cannot be mitigated by
statistics \citep[see their Appendix~A and
Figure~13,][]{Weisz2011ANGST}. Age errors in CMD analysis can be
smaller than in SEDs \citep[see][for a discussion]{Gallart2005}, but
our CMD-related results below are qualitatively unchanged by the error
model, so we will use the same smoothing filter for all comparisons.

In addition to resolution, there are substantial systematic uncertainties in
stellar population synthesis (SPS) models. 
\texttt{VESPA} analysis, using both \citet[hereafter BC03;][]{Bruzual2003} and the
\citet[hereafter M05;][]{Maraston2005} SPS models, is shown in plots. Those differences render
Poisson error on the mean (of order the point size), irrelevant for
all of the SDSS samples in Table~\ref{table:arch}. For
\citetalias{Weisz2011} dIrrs, statistical uncertainty is similar to
uncertainty between stellar evolution tracks (see their Figure~3). In
both cases, error estimates are only qualitative in the sense
that they illustrate the difference between models and not the
uncertainty in the model parameters.

\subsection{Consistency with SDSS SEDs} 
\label{sec:archresults}

\begin{figure*}[!t]
\begin{center}
\vspace{.5cm}
\includegraphics[height=0.55\textwidth]{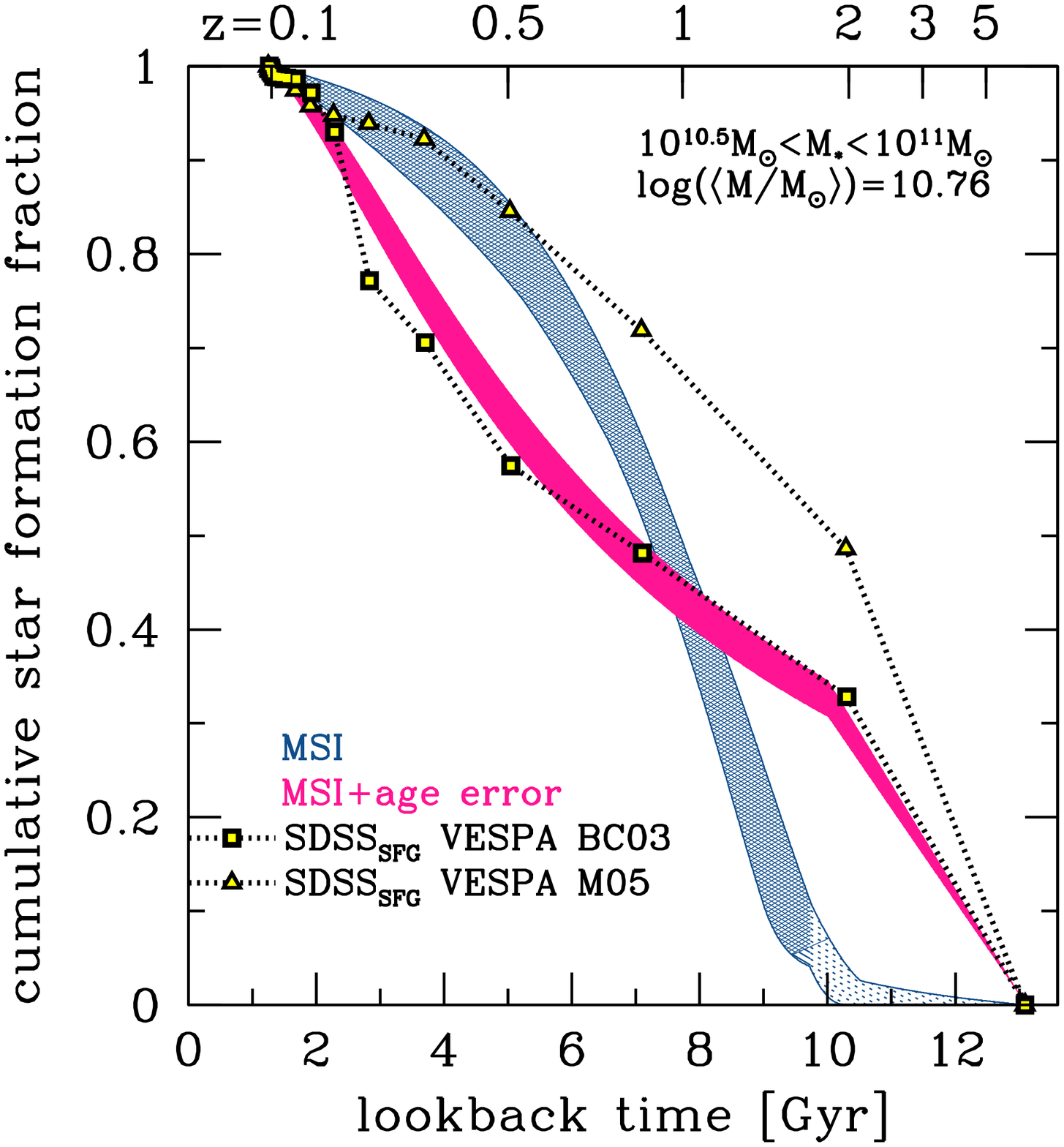}\hspace{0.0ex}
\includegraphics[height=0.55\textwidth]{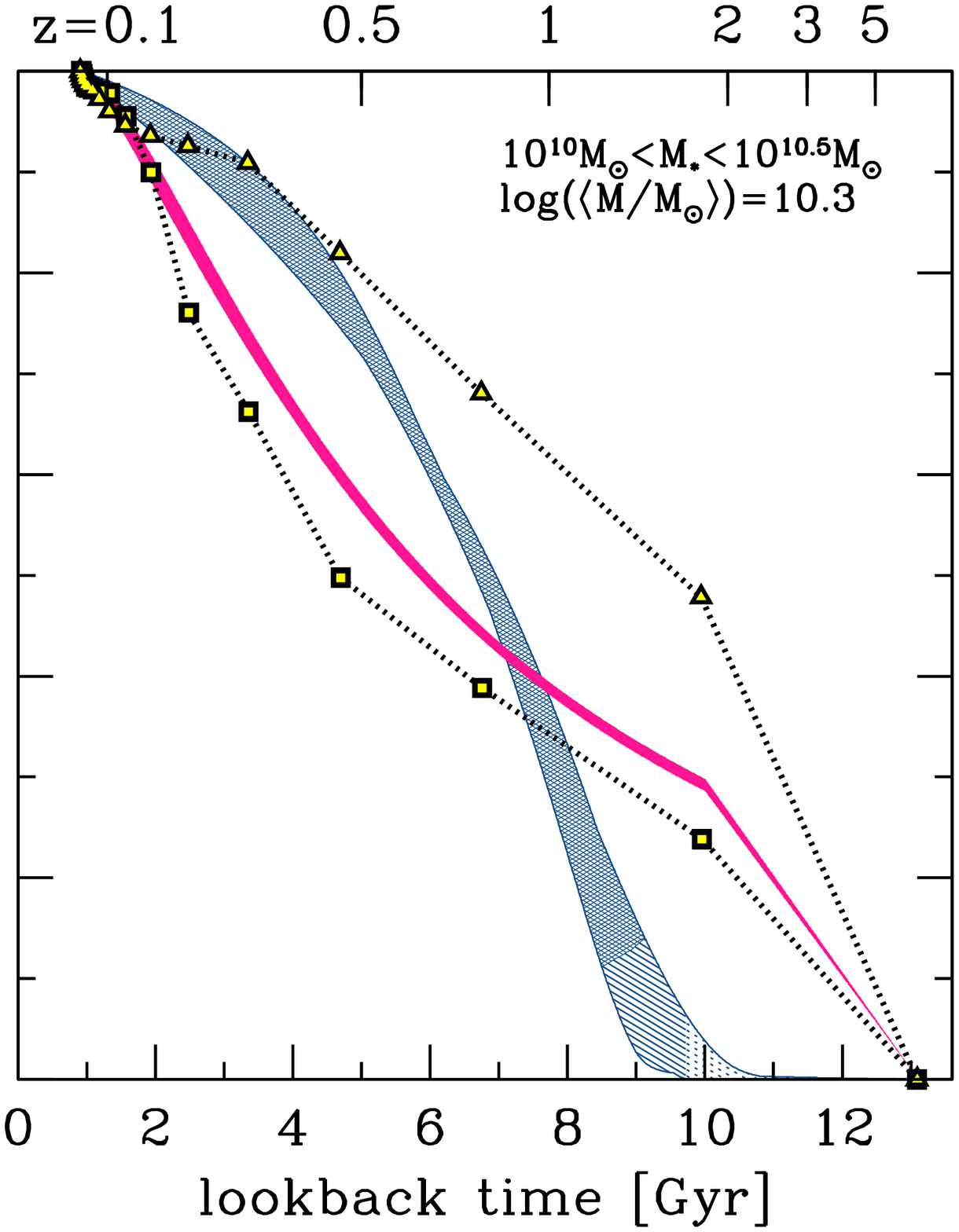}\hspace{0.0ex}\\
\includegraphics[height=0.55\textwidth]{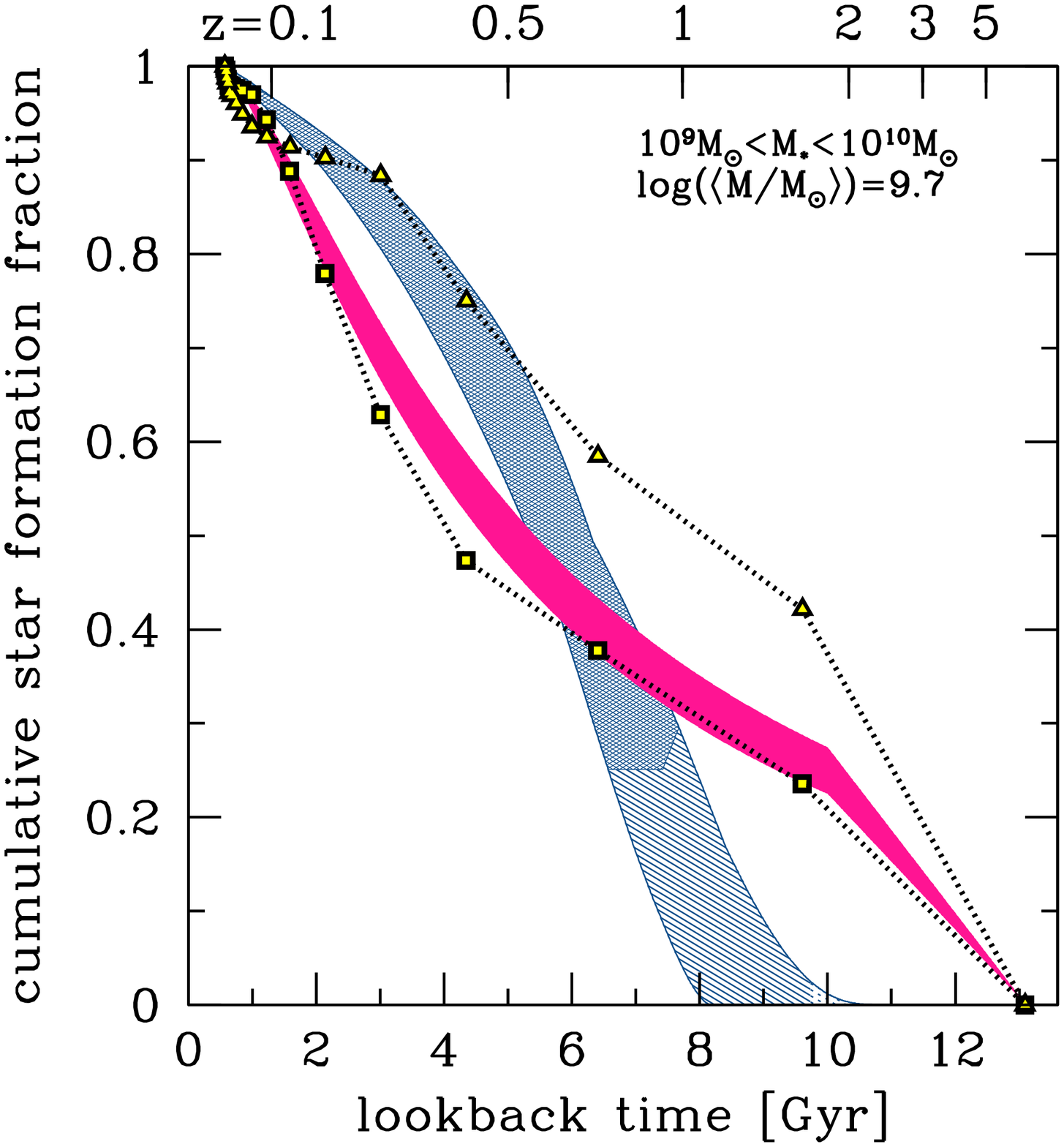}\hspace{0.0ex}
\includegraphics[height=0.55\textwidth]{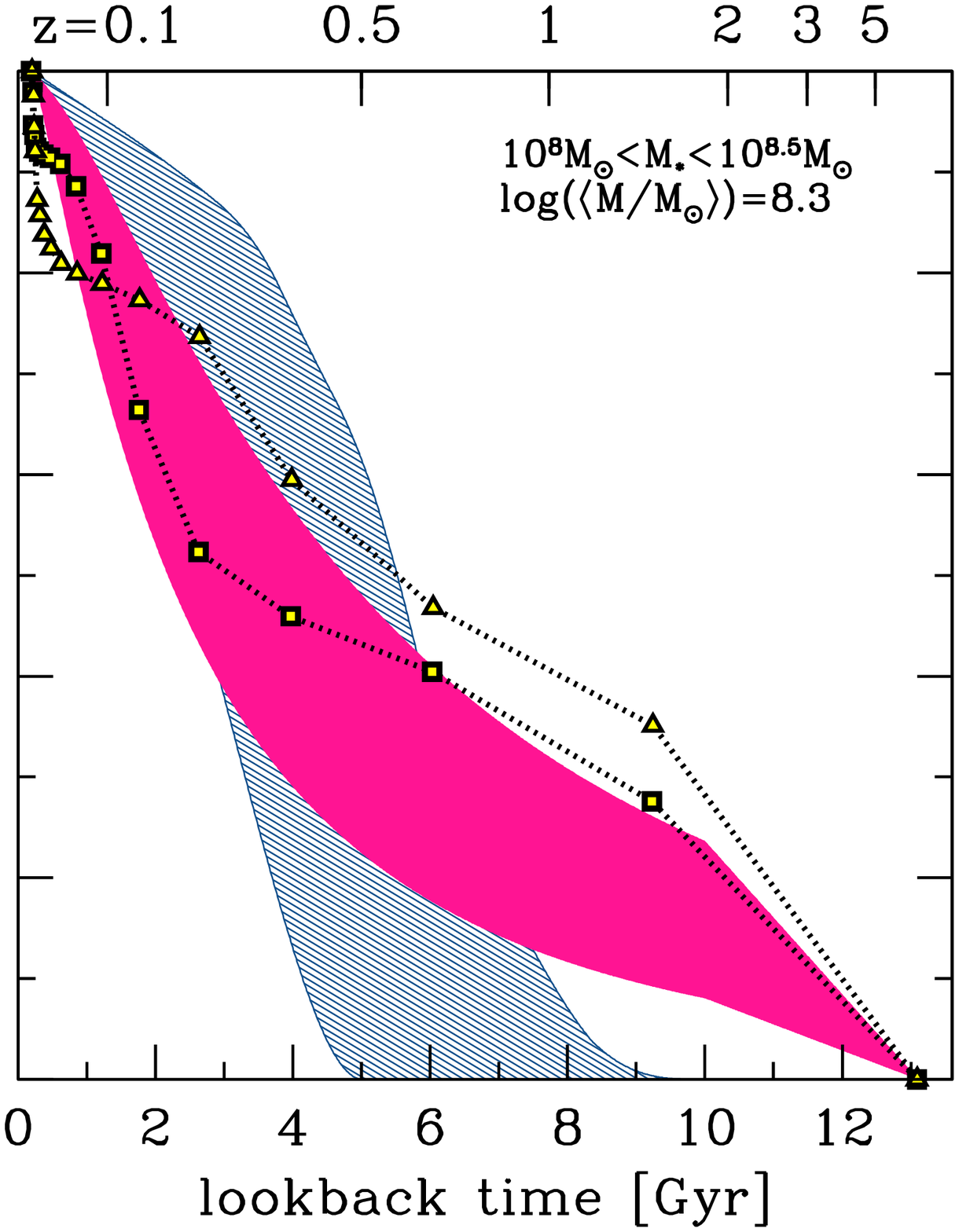}\hspace{0.0ex}
\end{center}
\caption{ Cumulative star formation from the SDSS SED modeling
  (points) and MSI (shaded regions) in four
  panels for different mass samples. 
\texttt{VESPA}'s SED analysis of
  SDSS is shown for the \citetalias{Maraston2005} (triangles), and
  \citetalias{Bruzual2003} (squares) population synthesis models. 
MSI (blue) was initialized with
  the same average mass and redshift as that of the SDSS samples
  and then smoothed to mimic age errors (magenta
  solid). The shading encompasses variation between
  \citetalias{Karim2011} and \citetalias{Oliver2010} SFR main sequence
  observations. Striping indicates the reliability of the results as
  described in Figure~\ref{fig:allgrowth}. 
}
\label{fig:MSI}
\vspace{.5cm}
\end{figure*}

Figure~\ref{fig:MSI} shows the average cumulative star formation
fraction\footnote{The cumulative star formation fraction is more
easily compared to observations than $M_\ast/M_{\ast0}$. However, the
two are similar and would be identical in the case of
fixed mass return fraction (i.e., constant $R(t)$). } for \texttt{VESPA} analyzed
SDSS SFGs and MSI, divided into panels for different stellar mass
bins. MSI models were initialized at the average mass and redshift of
the SED comparison sample (Table~\ref{table:arch}). Black points mark
\citetalias{Bruzual2003} (squares) and \citetalias{Maraston2005}
(circles) SPS analysis. These models show substantial differences that
are also visible in the (logarithmic) plots of previous studies
\citep[e.g.,][]{Panter2007,Tojeiro2009}, likely originating in their
different treatment of thermally-pulsing asymptotic giant branch (TP-AGB) stars. These
discrepancies remain the subject of active debate
\citep[e.g.,][]{Maraston2005, Conroy2010, Maraston2011, Kriek2010,
MacArthur2010}.

Focusing on the \texttt{VESPA} results, a cursory look suggests that SFGs appear to form stars
with an S-curve age distribution. In that case, average present-day
SFGs grew rapidly at high redshift ($z>2$) before biding their time at
intermediate redshifts -- while sSFRs were still $3-10$ times higher
than today. Finally, they surge again, forming stars rapidly over the
last $1-2\Gyr$.\footnote{These features were noted in past analysis
\citep{Tojeiro2009, Panter2007}, but attributed to TP-AGB stars
because \citetalias{Maraston2005} shifts some of the low redshift star
formation to earlier epochs (although the S-curve shape is still visible).}.
Such present-day SFGs would have occupied highly biased position in
the SFR main sequence, possibly leaving their mark in environmental or
structural trends.

However, such a naive interpretation of the data is unwarranted in
light of uncertainty in stellar population ages. This uncertainty
drives the peak of SFR to spread out, and the logarithmic nature of
variations in stellar populations naturally gives rise to that S-curve
in cumulative star formation. Specifically, intermediate to old
populations shift mass from a high(er) redshift SFR peak to low
redshift, while a tail of star formation extends to the beginning of
the universe (see figures in Appendix~\ref{sec:ageresolution} also).
Whatever the SED or CMD quality, the nature of this shape change
should be a generic part of properly regularized SED-based
analysis\footnote{Although the proper error model may vary depending
on features in SSP spectral templates.}. The impact on ``true'' MSI
predictions (blue) is a shift to SFHs that are well matched to the
\texttt{VESPA} \citetalias{Bruzual2003}\footnote{The match to
\citetalias{Bruzual2003} as opposed to \citetalias{Maraston2005}
should be interpreted cautiously. The SFHs reflect some combination of
the real inversion of flux from the SED, and the way noise is
interpreted in terms of the SSPs basis. A log-age error is expected,
but it may not capture all important artifacts in the flux inversion.} SED analysis
of SDSS (magenta) at all masses shown.

\subsection{Star Formation Trends and Local Dwarfs} \label{sec:dwarfs}

\begin{figure}[!t]
\begin{center}
\includegraphics[width=0.49\textwidth]{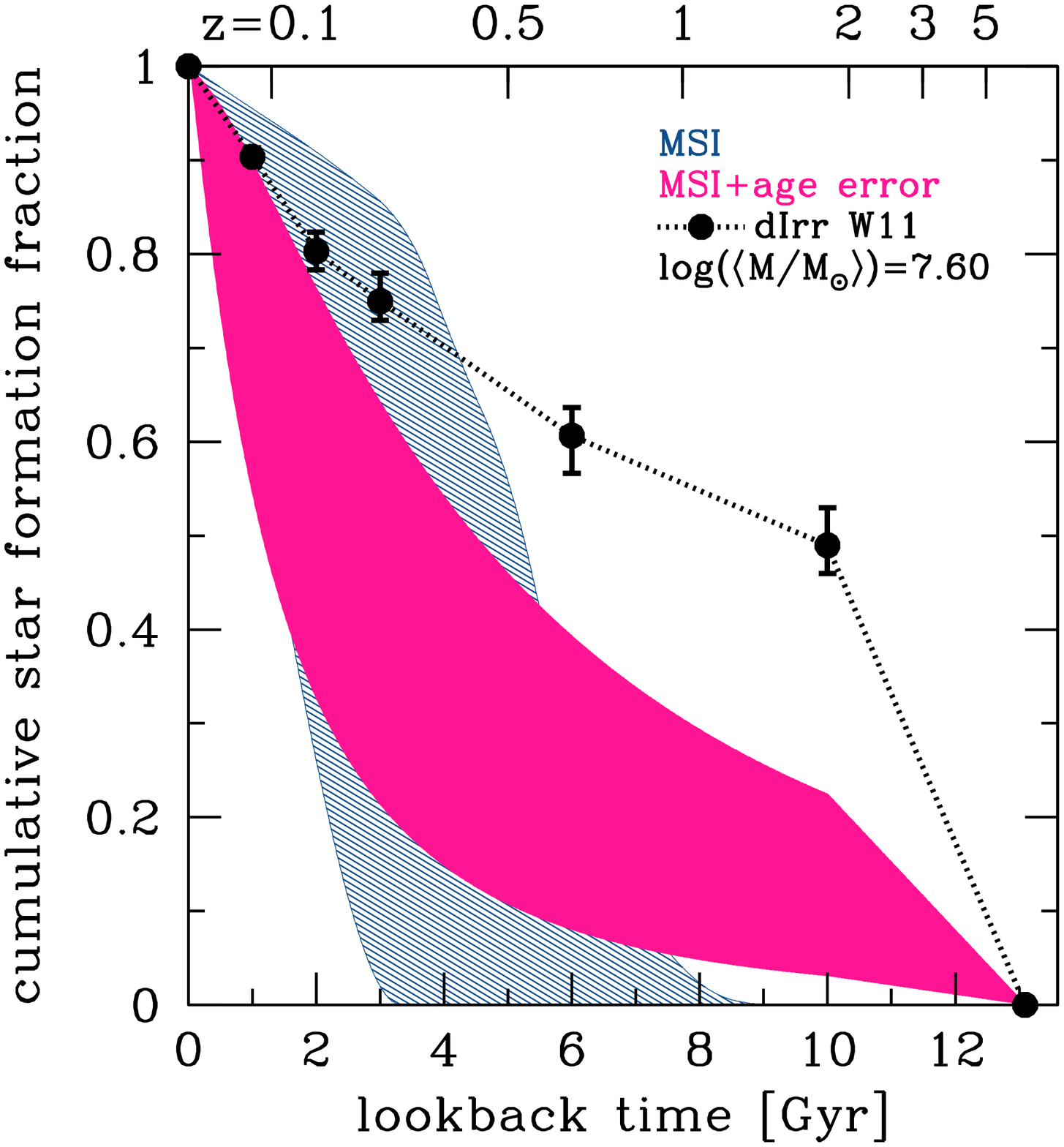}\hspace{0.0ex}
\end{center}
\caption{Same as Figure~\ref{fig:MSI}, but black points are
  \citetalias{Weisz2011} CMD-based results for dwarf irregulars
  (circles). Error bars show statistical uncertainty on the mean,
  which is similar to systematic variation between different stellar
  evolution models. MSI trends are extrapolated over this mass range.}
\label{fig:MSIdwarf}
\vspace{.5cm}
\end{figure}

For the \citetalias{Weisz2011} analysis of dIrr, on the other hand,
Figure~\ref{fig:MSIdwarf} shows significant differences from
extrapolation of MSI below the mass completeness range of SFR surveys.
Our simple model of age errors and those errors estimated in
\citetalias{Weisz2011} are unable to account for these differences. In
this regime, MSI predicts excessive downsizing such that dwarfs form
most of their mass below $z=1$, while the
\citetalias{Weisz2011} dIrr galaxies form most of their mass at $z>2$.

\begin{figure}[!t]
\begin{center}
 \includegraphics[width=0.49\textwidth]{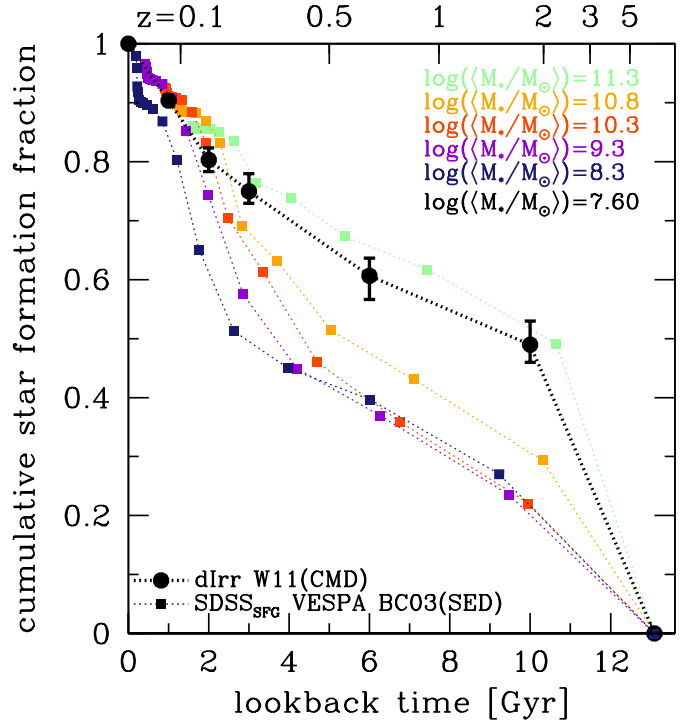}
 \end{center} 
\caption{ Cumulative star formation for SDSS SFGs from
 \texttt{VESPA} \citetalias{Bruzual2003} SED modeling in mass bins at
 $10^{8-8.5}$, $10^{9-9.5}$, $10^{10-10.5}$, $10^{10.5-11}$, and
 $10^{11-11.5}$, colored (squares) from lightest to darkest
 respectively, compared to the \citetalias{Weisz2011} dIrr sample
 (black circles). Each mass bin has been rescaled to the linearly
 interpolated cumulative star formation fraction of the dIrr sample at
 their epoch of observation. Star formation moves to lower redshift as
 mass decreases (downsizing), but the dIrrs form more like the most
 massive sample. }
\label{fig:dwarfcomp}
\vspace{.5cm}
\end{figure}

Figure~\ref{fig:dwarfcomp} demonstrates the extent of this shift in
the context of the \texttt{VESPA} \citetalias{Bruzual2003} SED analysis that
was matched by MSI trends. The figure shows SFHs in the range
$10^{11-11.5}M_\odot$, $10^{10-10.5}M_\odot$, $10^{9-9.5}M_\odot$, and
$10^{8-8.5}M_\odot$ (from lightest to darkest), plotted over the lower
mass \citetalias{Weisz2011} dIrrs (most with $M_\ast\lsim
10^8M_\odot$) that are scaled to the linearly interpolated dwarf mass
at $\langle z_{\rm obs}\rangle$ of the given sample. Archaeological
downsizing is apparent in the purely star-forming sample \citep[as
previously noted by, e.g.,][]{CidFernandes2005, Asari2007}.
Nevertheless, the \citetalias{Weisz2011} dIrr galaxies seem to reverse this trend:
they assemble more like galaxies of $\sim 10^{11}M_\odot$, and deviate
strongly from their $\sim10^{8}M_\odot$ siblings analyzed with
\texttt{VESPA}\footnote{To ensure that the downsizing trend we present at low
  masses was not purely the result of noisy SEDs, we also limited our
  sample to only galaxies with high quality spectra with average
  S/N$>20$ in the regions that are fit ($\langle \rm{S/N}\rangle=27$), and
  found identical results.} and trends in the SFR main sequence.

This is in line with a central conclusion of \citet{Weisz2011}, that
the formation of dwarfs in the local volume agrees with the cosmic SFH
(which is dominated by massive SFGs, and may be overestimated at
$z\gsim \zerr$), but here we have highlighted that these lowest
mass dwarf SFHs seem quite inconsistent with the cosmological
downsizing trends found in SFR observations and the SED-based fossil
record. A discussion of this shift, and its observational context
follows in \S\ref{sec:disharmony}.

\section{Discussion}\label{sec:discussion}


In section~\ref{sec:misfit} we showed that MSI indicates late
formation of typical SFGs ($z\lsim2$ and $\tlb\lsim10\Gyr$), then
compared the corresponding mass growth with the fossil record in
section~\ref{sec:archaeology}. In this section we place these MSI
results in a broader physical and observational context.

\subsection{A Simulation Context} \label{sec:discussion_sims}

Main sequence integration has no connection to galaxy dynamics, so it is interesting to
note that its implications align with the trends seen in increasingly successful cosmological simulations. 
In particular, the late-formation of SFGs suggested by MSI may
resolve the (potentially orthogonal) struggle to reproduce the
structure of late-type SFGs in simulations.

Current galaxy formation simulations typically convert too much low angular momentum gas
into massive bulges. The observed median bulge-to-total (B/T)
stellar mass ratio is $15\%$ in bright ($\ge10^{10}M_\odot$) local
late-type galaxies \citep[][]{Weinzirl2009, Gadotti2007}, and 
``bulgeless''\footnote{A bulgeless galaxy is defined as having S\`ersic index $n<2$ \citep[see,][for further discussion]{Graham2011}.} disk galaxies  are abundant in the local universe \citep[especially dominating at $<10^{10.5}M_\odot$][]{Kormendy2010,Fisher2011}. Such small bulges, embedded in
star-forming disks, are rarely reproduced in the current
generation of simulations of $M_{\ast0}\ge10^{10}M_\odot$ galaxies.

Importantly, whatever the physical or numerical mechanisms that
drive simulated bulge formation
\citep[e.g.,][]{Piontek2009,Brook2011,Governato2010}, state-of-the-art
simulations of SFGs consistently produce their excessively massive bulges at
$z\gsim2$ -- earlier than MSI indicates $\ssim15\%$ of stellar mass
should have assembled.
Specifically, \citet{Scannapieco2011} and \citet{Stinson2010} find
that their kinematic bulges comprise the majority of each one of their
17 galaxy's stellar masses; the \citet{Scannapieco2011}
bulges form at $z>2$, and the \citet[][Figure~14]{Stinson2010}
non-disk stars also form early. The higher resolution \citet{Brooks2011}
sample of 6 massive galaxies with $\BT\geq0.26$ ($\langle
\BT\rangle=0.44$) also form at least the majority of their mass at
$z>1$ \citep[A. Brooks private communication; also see,][and
discussion of sSFRs therein]{Brooks2011}. \citet{Guedes2011} simulated
a Milky Way-like disk galaxy with a (photometric) $\BT=0.26$,
again, with almost all of the bulge mass forming at $z>2$. Suppressed
star formation at $z>2$ is responsible for the reduced bulge fraction
in a parameter study of a very massive disk in \citet{Agertz2011}.
Finally, \citet{Brook2011} forms a $\BT=0.21$ disk with a SFH that
closely matches our fiducial SFH expectation for a $10^{10}M_\odot$
system (Figure~\ref{fig:allgrowth}). Delayed formation seems not
only achievable in simulations, but also may be important for reproducing
typical disk structures.

Intriguingly, two small dwarf galaxy simulations by
\citet{Governato2010} manage to form early without forming massive
bulges (one is fit by an exponential profile and one has bulge-to-disk ratio of $0.08$).
Such simulations may be important for understanding suppressed star
formation at low masses and high redshift for all galaxies. They may
also lead to a resolution of the unanticipated early growth of
\citetalias{Weisz2011} dwarfs (but see the next section).

\subsection{Resolved and High S/N Observations}
\label{sec:disharmony}

The abrupt shift in the CMD-based age distribution of
\citetalias{Weisz2011} dIrrs compared to SED- and MSI-based SFHs, could instead
highlight problems with MSI-, SED-, or CMD-based SFHs. The
handful of massive galaxies in \citet{Weisz2011ANGST} form even more
of their mass late, with the 0 most massive galaxies in their sample
($M_\ast>10^{7.75}M_\odot$) reported to have formed $64\pm0.6\%$ of
their stars at $z>2$. That is slightly more $z>2$ star formation than
in the lower mass dIrrs ($52^{+4}_{-3}\%$) and makes the nuances of
dwarf galaxy formation a less persuasive explanation for the
transition to early-forming dwarfs.

Another explanation is simply that the sample is small and biased by the local group overdensity. 
However, those same \citetalias{Weisz2011} dwarfs fit nicely into sSFR
trends when measured as part of the larger GALEX-HUGS11 sample of
\citet[][]{Lee2009UV}($\sSFR_{\rm K11}\left[10^{7.6}M_\odot,z=0\right]\approx\sSFR_{\rm LeeUV}\approx0.4\Gyr^{-1}$). 

Unfortunately, other analyses of resolved populations in SFG disks are limited in
number and ability to discern age differences, with too-large error
bars on mass formed in a given bin for statistically significant
inferences \citep[e.g.,][]{Williams2011}. Where any constraint on the
$z=2$ divide is reported, results are also mixed: NGC~300
\citep[][$2.4\times10^{10}M_\odot$]{Gogarten2010} formed $\sim70\%$ of
its stars at $z>2$, while M33
\citep[][]{Williams2009M33,Barker2007}, and NGC~2967
\citep{Williams2967} may have formed most of their mass at $z<2$.

\citet{MacArthur2009} and \citet{SanchezBlazquez2011} recently
performed full spectrum fits\footnote{At a minimum full spectrum
fitting with multiple SSPs is critical to age determinations of SFG
because of their extended SFHs. Single SSPs fits or Lick indices by
themselves are insufficient as detailed in \citep{MacArthur2009}} to
high S/N ($\geq50$) data in the central regions of nearby disk
galaxies. While these measurements target the inner disks and bulges
of galaxies, the results endorse strikingly old populations for the
bulk of stellar mass in galaxy disks. Of their collective 10 galaxies
($M_\ast\sim 10^{10}M_\odot$) with measurements to the disk
scale-length ($r_d$), six (I0239, N7495, N7490, N0173, N1358, and
N1365) appear to form most, in several cases more than $90\%$, of
their stars at $r_d$ before $z=2$. The blue regions in
Figure~\ref{fig:MSI} shows that galaxies of this mass would be
expected to form $10\%$ of their mass over the last $5\Gyr$. If such
galaxies form $90\%$ of their entire stellar mass at $z>2$, they would
have persisted as massive quenched disks for around $5\Gyr$.

If born out as representative, these results may demand either an
important recalibration of SPS models, or a shift in our understanding
of where present-day SFGs sit in the SFR main sequence (something not
apparent in studies of environment, or the structural main sequence,
see~\S\ref{sec:averaging}). More generally, such extreme examples
highlight the value of cross checks between the SFR main sequence and
fossil record analysis, especially of the sort recently undertaken by
\citet[][]{Wuyts2011}. Clearly, a larger number of high-S/N observations 
and robust examinations of disk ages will be crucial for a better
understanding of SFG growth and of the galaxy fossil record.

\section{Summary and Conclusions}\label{sec:conclusions}

We have derived the average growth of stellar mass in present-day
SFGs by simple integration of the consensus SFR main sequence under the assumption
that SFGs were always SFGs in the past (an approach that we term MSI). 
Our results regarding average stellar mass growth can be summarized as follows: 
\begin{itemize}
\item SFGs form late such that only $\sim15\%$ of the stellar mass in
  the progenitors of $5\times10^9M_\odot\lsim M_{*0}\lsim
  10^{11}M_\odot$ galaxies was in place before $z=1-2$
  (depending on mass and downsizing).
\item It follows that massive SFGs at $z>2$ are not expected to be progenitors of typical SFGs today.
\item The effect of mergers on MSI results was found to be negligible
  based on semiempirical merger trees generated in the MSI framework.
\item Similarly, there is no clear evidence that the way SFGs occupy the
  distribution of SFRs in the SFR$-M_*$ relation significantly 
  alter MSI results.
\item Accurate analytic approximations to average SFHs derived from
  MSI are presented in Appendix~\ref{sec:misfit_analytic}.
\end{itemize}
Moreover, the delayed formation of SFGs implied by our analysis \citep[but first noted in
the stage-$\tau$ models of][]{Noeske2007st} was found to be
consistent with observations of the evolution of stellar mass in the universe,
is also implied by halo-based semiempirical methods
\citep[][]{Conroy2009,Zheng2007}, and fits well in the context of
increasingly successful cosmological simulations.

When comparing MSI-based SFHs to those inferred from the fossil record, we found that, 
\begin{itemize}
\item Expected age uncertainties in SED-based analysis cause a characteristic shape distortion in derived SFHs. 
\item SED analysis of SDSS SFGs could be reconciled with MSI after accounting for these expected age uncertainties.
\item Local dwarf galaxies with $\langle M_* \rangle<10^8$, on the
other hand, were found to depart from MSI downsizing extrapolations,
with SFHs resembling those of $10^{11}M_\odot$ galaxies.
\end{itemize}
We stressed that the last point may reflect systematic errors in
population synthesis models rather than a physical transition. For a
handful of cases, CMD and high S/N SED analysis of more massive
galaxies suggests more early star formation than inferred from MSI. A
larger sample of SFGs is greatly needed and it is clear from our SDSS
comparison that a full understanding age uncertainties is crucial for
interpretation of that data. It will be interesting to see how
improved SED modeling and larger data sets can be accommodated in the
SFR main sequence.


\section*{Acknowledgments}
I am grateful to Andrey Kravtsov for numerous suggestions that
improved the clarity and scope of this paper, and to Rita Tojeiro for
help with VESPA. I would also like to thank Nick Gnedin, Hsiao-Wen
Chen, Fausto Cattaneo, Andrew Wetzel, Charlie Conroy, Mariska Kriek,
and Oscar Agertz for their thoughts and stimulating discussion. I am
indebted to many groups referenced in this study for making their data
public. This work was supported in part by the Kavli Institute for
Cosmological Physics at the University of Chicago through grants NSF
PHY-0114422, NSF PHY-0551142, AST-0507596 and AST-0708154 and an
endowment from the Kavli Foundation and its founder Fred Kavli. This
work made extensive use of the NASA Astrophysics Data System and
arXiv.org preprint server.

\appendix
\section{Analytic SFHs} \label{sec:misfit_analytic}
With a few approximations, accurate analytic formulae for average SFH
and stellar mass growth based on SFR observations can be derived. Since these may be
useful for tying chemical evolution or SPS models to SFR main sequence
measurements \citep[e.g.,][]{Wuyts2011} or for simulation comparisons,
they are provided below.

First, in the concordance cosmology, the expansion factor is linear as
a function of lookback time to $<5\%$ accuracy until
$\tlb\approx12.5\Gyr$ or $z\approx4$,
\begin{equation}\label{eq:approxz}
(1+z)\approx(1-\lambda_a \tlb)^{-1} ,
\end{equation}
where $\lambda_a=(1-a(t_{\rm lin}))/t_{\rm lin}$ and we take $t_{\rm
  lin}=12\Gyr$ so that $\lambda_a=0.064\Gyr^{-1}$ in our cosmology.
  Rewriting Eq.~\ref{eq:mstardot} in terms of the stellar mass
  fraction formed $f_\ast(t)\equiv\frac{M_\ast(t)}{M_{\ast0}}$, and
  with $R(t)\equiv\Re(t)/\Phi(t)$ being the fraction of the
  current SFR returned to the ISM by stellar mass loss, then, for a
  power law fit to the SFR main sequence (Eq.~\ref{eq:powSFRF}),
\begin{equation} \label{eq:fstardot}
\dot{f_\ast}(t) = \left[1-R(t)\right]\dot{f}_{0} {f_\ast(t)}^{1+\beta}(1-\lambda_a
t)^{-\alpha} ,
\end{equation}
where $\dot{f}_0 =\frac{A_{11}}{M_{\ast0}} \left(\frac{M_{\ast0}}{10^{11}M_\odot}\right)^{\beta+1}$. Taking $R(t)=0.45$ and integrating for $f_\ast$,
\begin{equation}\label{eq:fstar}
f_\ast(t) =  \left[  1 + \frac{\beta}{(\alpha-1)}\frac{\dot{f}_0\left[1-R(t)\right]}{\lambda_a}\left[(1+z)^{\alpha-1}
-1\right]\right]^{-\frac{1}{\beta}} .
\end{equation}
Then, plugging Eq.~\ref{eq:fstar} into Eq.~\ref{eq:fstardot}, analytic
SFHs, $\Phi(t)= M_{\ast0}\dot{f}_\ast(t)/\left[1-R(t)\right]$,
can be synthesized from a power law $\psi$:
\begin{equation} 
\Phi(t)=M_{\ast0}\dot{f}_{0} (1+z)^{\alpha} 
\left[  1 +
 \frac{\beta}{(\alpha-1)}\frac{\dot{f}_0\left[1-R(t)\right]}{\lambda_a}\left[(1+z)^{\alpha-1}
-1\right]\right]^{-\frac{1+\beta}{\beta} } .
\end{equation}
$\alpha$ and $\beta$ parameters can be found in
Table~\ref{table:fitdata} (under the ``pow'' form) and are commonly reported
with SFR sequence observations. Any broken power law in $\alpha$ or
$\beta$ is a simple extension (e.g., for a flattening of $\alpha$ at
$z\approx2$ or a flattening of $\beta$ in dwarfs).

\begin{figure*}[]
\includegraphics[width=0.49\textwidth]{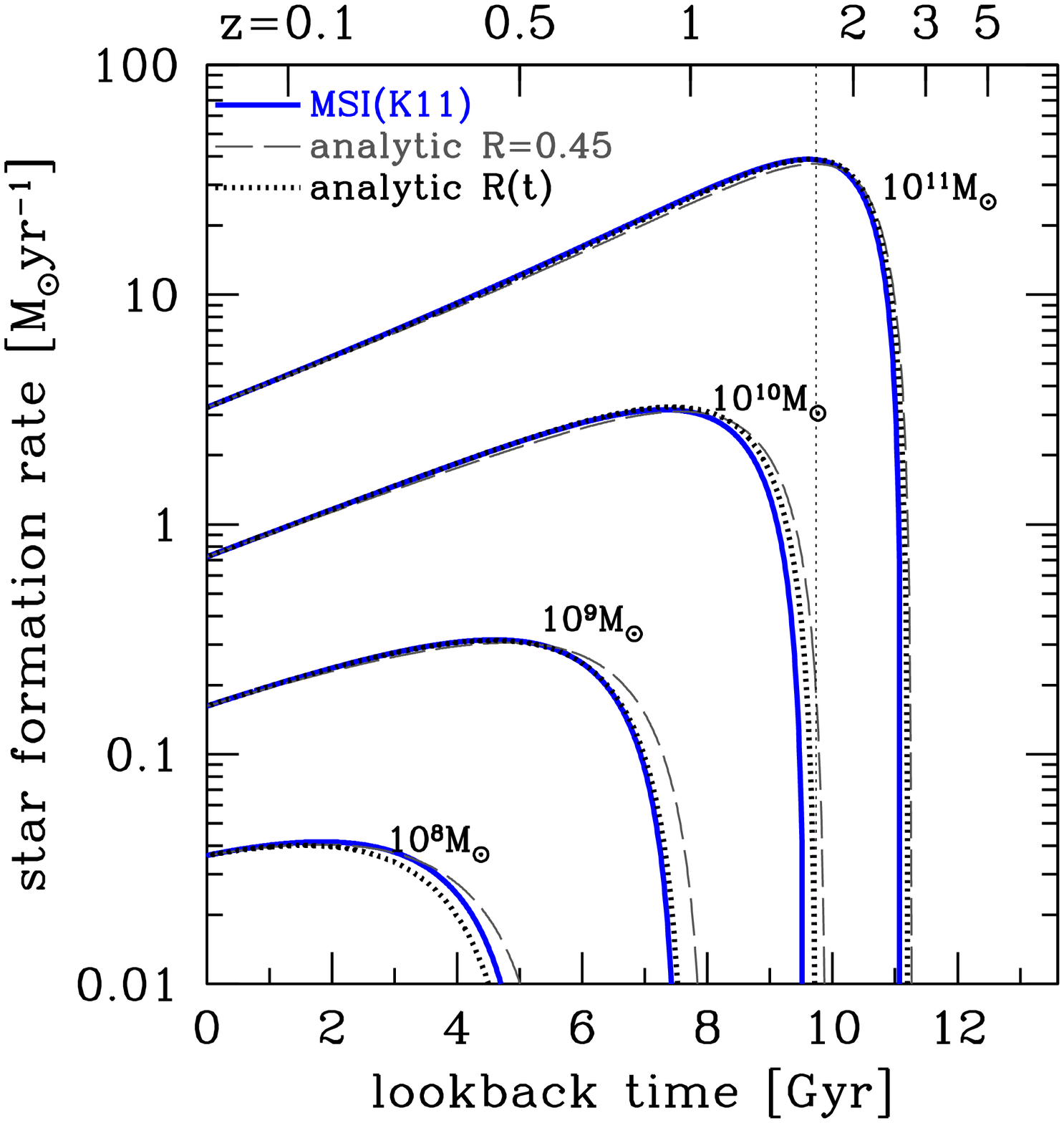}
\includegraphics[width=0.49\textwidth]{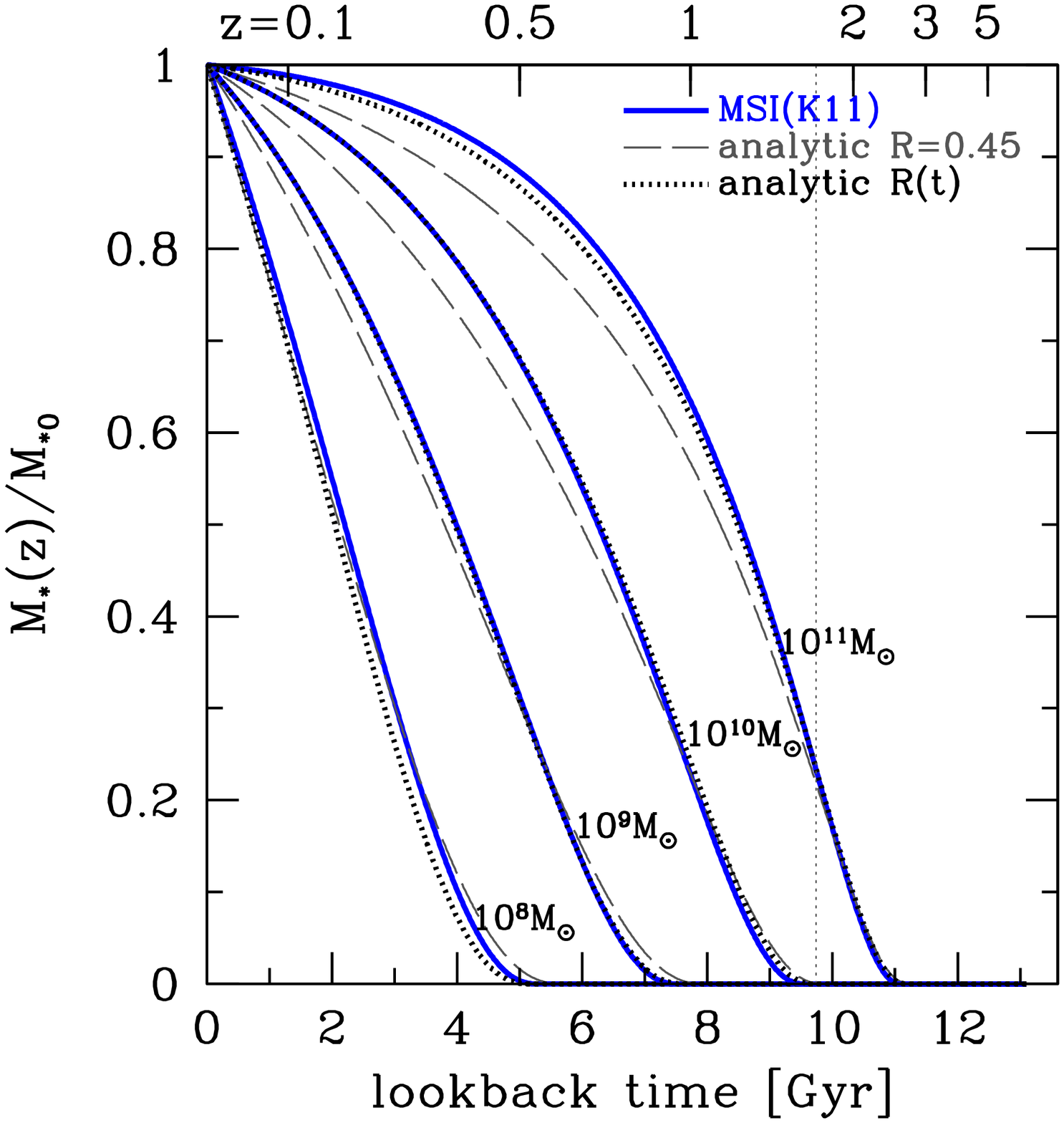}
\caption{SFHs (left) and mass growth (right) from MSI,
  in galaxies of $M_{\ast 0}=10^8M_\odot$, $10^9M_\odot$,
  $10^{10}M_\odot$ and $10^{11}M_\odot$. Calculations are shown for
  the fiducial SFR fit (\citetalias{Karim2011}) using the full MSI
  approach (blue lines), and analytic approximations with $\Re=0.45$
  (gray long-dashed), and $\Re(t)$ (black dotted; given by
  Eq.~\ref{eq:rapprox}). Vertical dotted lines are $\zerr$ from
  Figure~\ref{fig:csfhtime}. }
\label{fig:misfitanalytic}
\vspace{.5cm}
\end{figure*}

The dashed gray line in Figure~\ref{fig:misfitanalytic} shows analytic
SFHs and $f_\ast(t)$ for constant $R(t)=0.45$.
Results are in good agreement with the full MSI procedure, but
late mass growth is overestimated in massive galaxies
because mass loss from old stars is not properly accounted for when
assuming a constant return fraction \citep[see][]{Leitner2011}.

As a first order correction for massive galaxies, the return fraction can be decomposed
into a constant term from young stars, $R_0$, and a term related to
accumulated old stellar mass. The relative fraction of gas returned by
old stars is roughly proportional to a galaxy's age $(\tform-\tlb)$
(i.e., amount of stellar mass if star formation were constant since
$\tform$) calibrated to a reference return fraction, $R_{11}$, in a
galaxy of $10^{11}M_\odot$ with ${\tform}_{11}$. Then
\begin{equation}\label{eq:rapprox}
R(\tlb)=R_{0} \left( 1 + \frac{\left(\tform-\tlb\right)}{\tform}\frac{\tform}{{\tform}_{11}}\frac{R_{11}}{R_{0}} \right) , 
\end{equation}
assigning $f_{\ast\rm form}(\tform)\equiv0.1$: 
\begin{equation}
\lambda_a\tform=  1 - \left[  1 -
\frac{(\alpha-1)}{\beta} \frac{\lambda_a(1-f_{\ast\rm form}^{-\beta})}{\dot{f}_0\left[1-R(t)\right]}\right]^{-\frac{1}{\alpha-1}} .
\end{equation}
We calculate $\tform$ values using $R(t)=0.45$, and plug those
values into Eq,~\ref{eq:rapprox}, with $R_0=0.45$ and
$R_{11}=0.30$. Inserting the full return rate from
Eq.~\ref{eq:rapprox} into Eq.~\ref{eq:fstar} results in the $\Phi(t)$
and $f_\ast(t)$ values plotted (dotted lines) over the full MSI
models (solid lines) in Figure~\ref{fig:misfitanalytic}. The
$f_{\ast}$ and SFH approximations are accurate to a couple of percent
of the final stellar mass (i.e., $20\%$ error when the galaxy is
$10\%$ of $M_{\ast0}$). The same correction are equally accurate
for the \citetalias{Oliver2010} parameterization.

\section{Notes on Age Resolution} \label{sec:ageresolution}
\begin{figure*}[!t]
\begin{center}
\vspace{.5cm}
\includegraphics[width=0.49\textwidth]{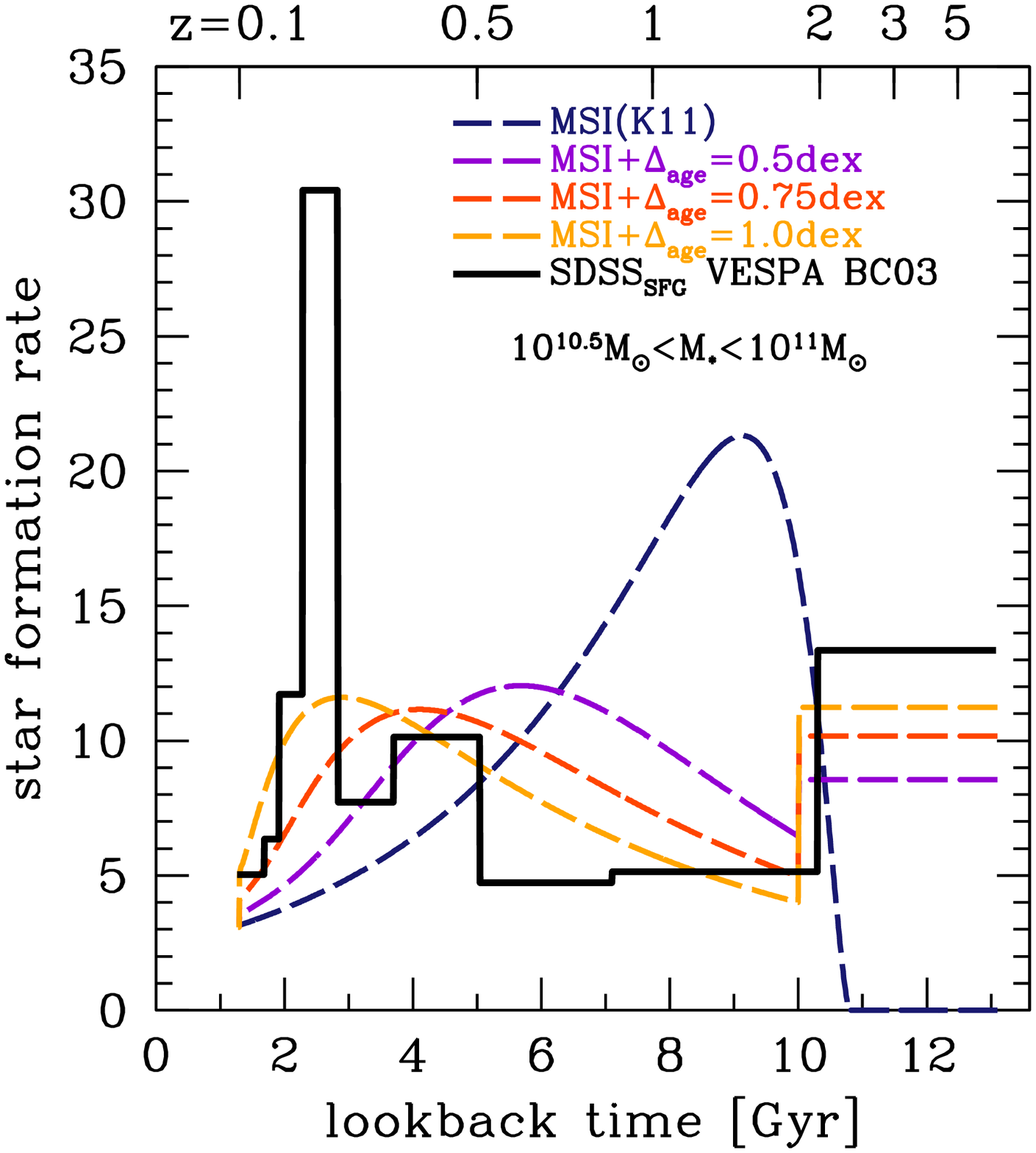}
\includegraphics[width=0.49\textwidth]{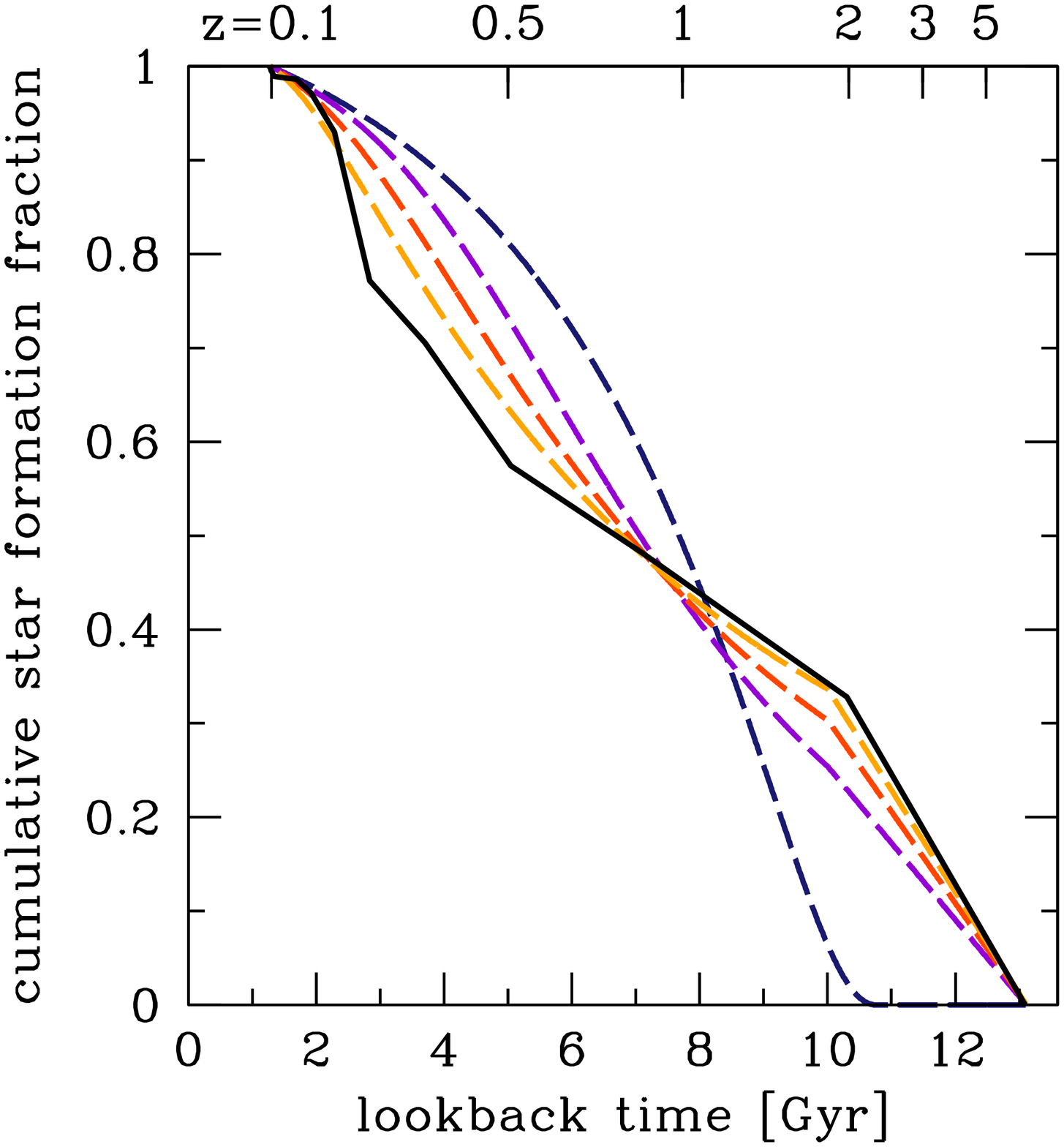}
\end{center}
\caption{ SFH ({\bf left}) and cumulative star formation fraction ({\bf right}) for SDSS SFGs of $10^{10.5-11}M_\odot$ from \texttt{VESPA} analyzed with \citet{Bruzual2003} (solid black), compared to MSI (dashed) with varied age resolution (see legend). MSI starts with the same average mass and look back time as the SDSS sample. The first 7 \texttt{VESPA} age-bins were combined to remove artifacts. 
}
\label{fig:spserr}
\vspace{.5cm}
\end{figure*}

In \S\ref{sec:archresults} we show that SFHs based on
the fossil record can be misleading, particularly about early galaxy
growth. The root of this claim is the consensus view that the
maximum reliable precision of age determinations in SED-analysis is
low \citep[$>0.5\dex$][]{Ocvirk2006,Walcher2011,CidFernandes2007}. Our
solution is a similarly uncontroversial smoothing of the SFH to mimic
age errors. The result, however, is a notable change in the shape of
SFHs that brings \texttt{VESPA} results into line with cosmological SFR trends
(i.e., the SFR main sequence). In light of the importance of noise
modeling to this result, a discussion of age errors is prudent. For
guidance, we rely on \citet{Ocvirk2006} who carried out a thorough
simulation campaign to quantify the noise properties of SSPs and full
spectrum fitting in PEGASE-HR \citep{LeBorgne2004} with implications
for interpretation of any optical SED inversion\footnote{\citet{Ocvirk2006} conclusions were reached over the
  slightly smaller spectral range, from $4000-6800$\AA~compared to
  \texttt{VESPA}'s $38000-7800$\AA~(including masked emission lines)
  .}. This section provides a short summary of the issue, and a
discussion of how it pertains to \texttt{VESPA} results.

Modeling the intrinsic SEDs of galaxies involves the conflicting goals of (1) finding a
stellar population (metallicity and extinction) that
generates a good fit to SED, and (2) not letting that population be shaped by 
noise in SEDs. Even fitting in the absence of noise is
not clear cut because stellar evolution models do not match all of the
observed features in galaxy spectra (see e.g., \citealp{Tojeiro2007};
\citealp{Panter2007}; \citealp{CidFernandes2005};
\citealp{Conroy2010}; \citealp{Walcher2011}) and the calibration of
important pieces of stellar evolution (e.g., TP-AGB stars at long
wavelengths: \citealp{Maraston2005, Conroy2010, Maraston2011,
  Kriek2010} ; stellar abundance: \citealp{Panter2007, MacArthur2009};
IMF, etc.) are ongoing. \citet{Ocvirk2006} found that noise calibrated
into the model PEGASE-HR spectra themselves place a fundamental floor
of $\sim 0.4 \dex$ on age recovery for even the highest S/N spectra.
They also found that noise in SEDs typical of SDSS spectra
(S/N$\approx20$ per $\sim3$\AA~pixel in the unmasked \texttt{VESPA} range)
renders differences between simple SSPs unrecoverable within
$\sim0.5\dex$. Any smaller-scale time information derived from SEDs
must come from noise, improved modeling, or the use of a
larger spectral range.
Moreover, these resolutions are further broadened when
considering non-SSPs with uncertain metallicities, extinction and
physical effects \citep[e.g.,][]{Wuyts2009}. Regularization
techniques can be used to assign resolution to solutions so that they
are consistent with these noise limitations and primarily responsive to signal.

\texttt{VESPA} attempts to regularize the stellar population solution using the
general procedure of \citet[][STECMAP]{Ocvirk2006}, but adaptively
refined logarithmic age bins for basis vectors. Given a solution basis
of age bins, metallicities and extinction, \texttt{VESPA} estimates the
contribution of noise to the coefficients (parameters) that multiply
the solution basis. The number of parameters used to describe the
solution should be $\leq\kappa$, the number of coefficients that are
not noise dominated \citep[see][\S2.2.2 ]{Tojeiro2007}. Then, to
select an appropriate age basis, the bins that contribute the most to
the total flux in the model are iteratively refined until they reach
the maximum $\Dage\approx0.2\dex$ resolution. $\kappa$ is
recalculated at each iteration. Finally, the chosen solution is the
most refined solution that has fewer non-zero flux age bins (and other
model parameters) than $\kappa$.

Contrary to the results of \citep{Ocvirk2006} that $\Dage<0.4\dex$ is not achievable, 
\texttt{VESPA} regularly refines to its highest resolution $\Dage\approx0.2\dex$. A full exploration of this
discrepancy is well beyond the scope of this paper, but one plausible
explanation is that, since bins containing no flux do not count toward
the parameter limit, $\kappa$, \texttt{VESPA} favors higher frequency solution
variations than can be robustly extracted (R. Tojeiro, private
communications). In contrast, STECMAP preferentially weights smooth
basis vectors for their solutions using well tested algorithms to
eliminate artifacts. Although it is not clear that the smoothed
STECMAP basis is ideal (as noted by \citealp{Ocvirk2006}), the \texttt{VESPA}
basis appears to be to aggressive. Ill conditioning of the inversion
means there is no way to know exactly how noise-modified light gets
interpreted into a stellar age spread.

Artifacts are to be expected but, supported by covariance primarily
between adjacent bins in \texttt{VESPA} tests, we assume that \texttt{VESPA} will mimic
the STECMAP regularization. That regularization favors
small gradients in $\log(t_{\rm age})$, finds little bias in $\langle \log(t_{\rm age}) \rangle$, and
almost constant logarithmic resolution $\pm\Dage/2$ about any median log-age value. In this case a
log-age Gaussian convolution with constant full width half max ($\Dage$) is the obvious
choice for a noise-model. The $1\dex$ resolution, as noted in
\S\ref{sec:noisyarch}, is motivated both by the \citep{Tojeiro2007}
covariance between adjacent bins and the typical mass-weighted size of the
bins. The convolution preserves $\langle \log(t_{\rm age})\rangle$ and biases $\langle t_{\rm age}\rangle$, depending on how  the $t_{\rm age}$ distribution is limited by cosmology.

Figure~\ref{fig:spserr} demonstrates that the $\Dage$ size for massive
SFGs, does not qualitatively alter our results. The figure shows the
SFH and cumulative star formation recovered by \texttt{VESPA} for galaxies of
$10^{10-10.5}M_\odot$ using \citet{Bruzual2003} analysis binned and
averaged at the highest age resolution (black solid line), compared to
MSI models (dashed) smoothed in log-age with $\Dage = 0, 0.5, 0.75$
and $1\dex$. The first seven \texttt{VESPA} bins were combined to hide
(dramatic) artifacts in these bins The S-curve noted in
\S\ref{sec:archresults} grows (right) with increasing $\Dage$ and,
correspondingly, the SFH peak flattens (left).

\bibliographystyle{apj}
\bibliography{lit}
\end{document}